\documentstyle{article}

\newcounter{compact}
\newcommand{\iscompact}{\setcounter{compact}{1}}

\newcommand{\ifcompact}{\ifnum\thecompact>0}
\newcommand{\ifnotcompact}{\ifnum\thecompact=0}

\iscompact
\newcommand{\doublespace}{}

\input{psfig}
\newcommand{\psfigure}[2]{\psfig{figure=#1,width=#2}}

\makeatletter  

\ifcompact
	\def\sectionsize{\normalsize}
\else
	\def\sectionsize{\large}
	\fi

\def\section{\@startsection {section}{1}{\z@}
	{-2.5ex plus -1ex minus -.2ex}      
	{1.2ex  plus .2ex}                  
	{\sectionsize\sf}}
\def\subsection{\@startsection{subsection}{2}{\z@}
	{-2.5ex plus -1ex minus -.2ex}
	{0.2ex plus .2ex}
	{\sectionsize\sf}}
\def\subsubsection{\@startsection{subsubsection}{3}{\z@}
	{-2.5ex plus -1ex minus -.2ex}
	{0.2ex plus .2ex}
	{\bf}}
\def\paragraph{\@startsection {paragraph}{4}{\z@}
	{1ex}     
	{-1ex}
	{\bf}}
\def\subparagraph{\@startsection{subparagraph}{4}{\parindent}
	{1ex}
	{-1ex}
	{\bf}}

\long\def\@makecaption#1#2{
 \vskip 10pt
 #2
}

\newcommand{\kaption}[1]{\caption[#1]{{\bf \small FIG.~\thefigure.}
{\small #1}}}

\def\thebibliography#1{\section*{REFERENCES}\list
 {\arabic{enumi}.}{\settowidth\labelwidth{[#1]}\leftmargin\labelwidth
 \advance\leftmargin\labelsep
 \itemsep 0ex
 \parsep 0ex
 \usecounter{enumi}}
 \def\newblock{\hskip .11em plus .33em minus .07em}
 \sloppy\clubpenalty4000\widowpenalty4000
 \sfcode`\.=1000\relax}

\def\@cite#1#2{#1\if@tempswa, #2\fi}

\def\onlinecite{%
\@ifnextchar[{\@tempswatrue\@CITEX}{\@tempswafalse\@CITEX[]}}

\def\@bylinecite{%
\@ifnextchar[{\@tempswatrue\@citex}{\@tempswafalse\@citex[]}%
}

\def\@CITEX[#1]#2{%
  \if@filesw\immediate\write\@auxout{\string\citation{#2}}\fi
  [\@cite{\@collapse{#2}}{#1}]%
}%

\def\@citex[#1]#2{%
\if@filesw\immediate\write\@auxout{\string\citation{#2}}\fi
\leavevmode\unskip\nobreak$^{\scriptstyle\@cite{\@collapse{#2}}{#1}}$}

\def\@collapse#1{%
{
\sfcode`,=2000\relax
\dimen0=\the\fontdimen2\textfont0\relax
\xspaceskip = .6\dimen0 plus.3\dimen0 minus.3\dimen0\relax
\let\@temp\relax
\@tempcntb\@MM
\def\@citea{}%
\@for \@citeb:=#1\do{%
  \@ifundefined{b@\@citeb}%
    {\@temp\@citea{\bf ?}%
     \@tempcntb\@MM\def\@citea{,}\let\@temp\relax%
     \@warning{Citation `\@citeb ' on page \thepage\space undefined}%
    }
    {\@tempcnta\@tempcntb \advance\@tempcnta\@ne
     \edef\MyTemp{\csname b@\@citeb\endcsname}%
     \def\@tempa{\@temptokena=\bgroup}
     \afterassignment\@tempa 
     \@tempcntb=0\MyTemp\relax}%
     \ifnum\@tempcntb=0\relax
       \@tempcntb=\@MM
       \@citea\@stripdollars{\MyTemp}%
       \let\@temp = \relax
       \def\@citea{,}%
     \else 
       \edef\@tempd{\number\@tempcntb}%
       \ifnum\@tempcnta=\@tempcntb%
          \ifx\@temp\relax
             \edef\@temp{\@citea\@tempd}%
          \else%
             \edef\@temp{\hbox{--}\@tempd}%
          \fi%
       \else%
         \@temp\@citea\@tempd
         \let\@temp\relax
       \fi
     \def\@citea{, }%
     \fi%
    }
}
\@temp 
}
}

\let\cite\onlinecite

\newcommand{\ignore}[1]{}
\newcounter{comment}
\newcommand{\tcomment}[1]{%
{\tt [* Comment: }#1{\tt *]}%
}
\newcommand{\comment}[1]{%
\addtocounter{comment}{1}%
{\tt [*\thecomment*]}%
\renewcommand{\thefootnote}{}%
\footnotetext{%
\renewcommand{\baselinestretch}{1.0}\normalsize\footnotesize
{\tt [** Comment~\thecomment:} #1 {\tt **]}}%
\renewcommand{\thefootnote}{\arabic{footnote}}%
}

\newcommand{\be}{\begin{equation}}
\newcommand{\ee}{\end{equation}}
\newcommand{\bea}{\begin{eqnarray}}
\newcommand{\eea}{\end{eqnarray}}
\newcommand{\bc}{\begin{center}}
\newcommand{\ec}{\end{center}}

\newcommand{\sect}[1]{Sect.~\ref{sec-#1}}
\newcommand{\fig}[1]{Fig.~\ref{fig-#1}}
\newcommand{\eq}[1]{Eq.~(\ref{eq-#1})}

\def\etal{~{\it et al.}}

\makeatother   

\ifnotcompact
	\renewcommand{\kaption}[1]{\caption[#1]%
	{\hfill\fbox{FIG.~\thefigure}\hfill}}
	\renewcommand{\psfigure}[2]{}
\fi

\ifcompact
\else
\fi

\begin{document}


\renewcommand{\tcomment}[1]{}
\renewcommand{\comment}[1]{}


\newcommand{\abclabel}{
    \centerline{
	    \hspace*{\fill} {\bf (a)} \hspace*{\fill}
        \hspace*{\fill} {\bf (b)} \hspace*{\fill}
        \hspace*{\fill} {\bf (c)} \hspace*{\fill}}
}
\newcommand{\ablabel}{
    \centerline{
	    \hspace*{\fill} {\bf (a)} \hspace*{\fill}
	    \hspace*{\fill} {\bf (b)} \hspace*{\fill}}
}

\def\whatthisis{article}


\newcommand{\thetitle}{
	\begin{flushright}
	\tt {\sl Chaos} {\bf 4} (December, 1993)
	\end{flushright}
	\vspace*{.1in}
	\begin{flushleft}
		{\LARGE\bf
		Don't bleach chaotic data\\[3ex]
		}
	\end{flushleft}
	\begin{quote}
		{\large\sf
		James Theiler\\[.5ex]
		{\it Center for Nonlinear Studies and Theoretical Division,
	Los Alamos National Laboratory, Los~Alamos, NM~~87545; {\rm and}
	Santa Fe Institute, 1660 Old Pecos Trail, Santa Fe, NM~~87501.}\\[2ex]
		Stephen Eubank\\[.5ex]
		{\it Center for Nonlinear Studies and Theoretical Division,
	Los Alamos National Laboratory, Los~Alamos, NM~~87545;
	Santa Fe Institute, 1660 Old Pecos Trail, Santa Fe, NM~~87501; {\rm and}
	Prediction Company, 320 Aztec Street, Santa Fe, NM~~87501.}\\[2ex]
		}
		(Received 10 July 1992; accepted for publication 11 August 1993)
	\end{quote}
}
\newcommand{\theabstract}{
	A common first step in time series signal analysis involves
digitally filtering the data to remove linear correlations.  The
residual data is spectrally white (it is ``bleached''), but in
principle retains the nonlinear structure of the original time series.
It is well known that simple linear autocorrelation can give rise to
spurious results in algorithms for estimating nonlinear invariants,
such as fractal dimension and Lyapunov exponents.  In theory, bleached
data avoids these pitfalls.  But in practice, bleaching obscures the
underlying deterministic structure of a low-dimensional chaotic
process.  This appears to be a property of the chaos itself, since
nonchaotic data are not similarly affected.  The adverse effects of
bleaching are demonstrated in a series of numerical experiments on
known chaotic data.  Some theoretical aspects are also discussed.
}


\thispagestyle{empty}

\ifnotcompact
	\thetitle
	\bc {\bf ABSTRACT} \ec
	\begin{quote}
	\doublespace
	\large
	\theabstract
	\end{quote}

	\vfill
	\ignore%
	{
	\small
	\renewcommand{\section}[2]{}
	\renewcommand{\numberline}[2]{\hspace*{1em}{\makebox[2em][l]{#1} #2}}
	\renewcommand{\contentsline}[3]{#2 \\}
	{\noindent\bf Contents:\\}
	\tableofcontents
	}

	\vfill
	\noindent\rule{2in}{1pt}
	\begin{flushleft}
	\end{flushleft}
\fi

\newpage


\ifcompact
	\pagenumbering{arabic} 
	\twocolumn[\thetitle
	\begin{quote}\noindent{\theabstract}\end{quote}]
	\small
	\columnsep 0.35in
	\sloppy
	\flushbottom
	\def\halffigsize{1.4in}
	\def\fullfigsize{3in}

\else
	\renewcommand{\kaption}[1]{\caption[#1]{\hfill\fbox{Fig. \thefigure}\hfill}}
	\renewcommand{\psfig}[1]{}
	\renewcommand{\abclabel}{}
	\renewcommand{\ablabel}{}
	\doublespace
	\large
\fi

\section{INTRODUCTION}

Much of the current interest in nonlinear signal processing arises not
so much as an extension of linear analysis, but from the recognition
that an entirely new idea -- chaos -- will play a significant role.
In some cases, this entirely new idea has led to entirely new
techniques for time series analysis.  These have provided
experimentalists with new ways to understand the implications of their
data, though the limitations of these new technologies have not always
been understood or well appreciated.  In other cases, chaos has shed
new light on the interpretation of conventional time series analysis
tools (for instance, by providing a deterministic explanation for
broadband spectra).  Our intent here is to investigate the limitations
of one of these conventional tools in the context of chaotic time
series.

Bleaching, or ``pre-whitening,'' is the process of linearly filtering
time series data to remove autocorrelation --- that is, to make the
power spectrum more nearly flat, or ``white.''  As a first step in
time series analysis, it is a time-honored practice among
statisticians~\cite{McLeod83,Keenan85,Tsay86,Tong90} and
statistics-minded
economists~\cite{Brock86,Brock86c,Brock91d,Hsieh89,Hsieh91,LeeTH92}.
Even the classic treatise of Blackman and Tukey~\cite{Blackman59}
recommends ``preemphasis'' of a signal to make the spectrum ``more
nearly constant.''  It is an initially attractive procedure because it
eliminates autocorrelation, which is one of the major sources of
artifact in nonlinear time series
analysis~\cite{Theiler86,Osborne89,Provenzale91,Theiler91,Rapp93}.
Further, since bleaching is accomplished with a finite order
non-recursive (or finite-impulse-response, or FIR) filter, it can be
proven that the nonlinear properties (such as dimension and Lyapunov
exponent) remain
invariant~\cite{Brock86c,Sauer91,Broomhead92,Sauer93,Isabelle92,Pecora93}.

However, this theoretical invariance does not always carry over to
practical data analysis.  It has long been known that recursive (or
infinite-impulse-response, or IIR) filters can --- in practice
and in principle --- change the character of a nonlinear process, as
inferred from its time series~\cite{Badii88,Mitschke88}.
Mitschke~\cite{Mitschke90} suggested that {\em acausal} IIR filters might
be less destructive, though others~\cite{Rapp93,Pecora93} have shown
that these too can change the nonlinear invariants.
Insofar as FIR filters approximate
IIR filters, their effects can be similarly detrimental: a graphic
demonstration is provided in Ref.~\cite{Sauer93}.  In an
earlier paper~\cite{Theiler91b-x}, we briefly noted that bleaching with
very high order linear filters can degrade evidence for nonlinearity
in a time series.  In this paper, that observation is extended.
Even when the bleaching is
constrained to relatively low order (by the Akaike criterion, for
instance), and even for tasks other than detecting nonlinear structure,
we find that the effect of bleaching on chaotic data can be detrimental.
On the other hand, bleaching {\em non}chaotic data does {\em not}
have such a negative effect.

After introducing the bleaching process in \sect{bleach}, the effect
of bleaching on chaotic data is demonstrated numerically, first by
looking at the problem of nonlinear prediction (in
\sect{modeling}), then by comparing residual-based to
surrogate data approaches for detecting nonlinearity in time series (in
\sect{test-nonlin}).  These numerical results lead us to argue
against pre-whitening chaotic data; however, in
\sect{misc}, this view is tempered by showing that some linear
prefiltering can still be advantageous.  The emphasis in this paper
is on numerical results, but in \sect{theory} some theoretical issues
are discussed: the limit of infinite data with infinite order
filtering; and the relation of filtering to the more familar problem
of ``optimal'' embedding.

\section{BLEACHING}
\label{sec-bleach}
Given a time
series $x_t$, the best linear predictor $\hat x_t$ is given by the model
\be
	\hat x_t = a_o + \sum_{k=1}^q a_k x_{t-k}
\ee
for which $\hat x_t$ most closely approximates $x_t$ in the least-squares
sense.  The residuals (also called ``innovations'' or ``disturbances'')
$e_t=x_t-\hat x_t$ measure how much of the
original time series is not linearly predictable from the past.
That is,
\be
	e_t = x_t - \left[ a_o + \sum_{k=1}^q a_k x_{t-k} \right],
	\label{eq-bleach}
\ee
where $q$ is the order of the model, and the linear coefficients
$a_k$ are obtained by a least-squares fit which minimizes the
variance of the residuals.

A result from the theory of linear time series analysis states that in
the large $q$ limit, the residuals $e_t$ obtained by subtracting from
$x_t$ the best linear predictor $\hat x_t$ will be uncorrelated; that
is, the residuals $e_t$ are spectrally white (the Appendix outlines an
informal proof).

\begin{figure}[bhtp]
\ \centerline{\hbadness=20000\hbox{
\psfigure{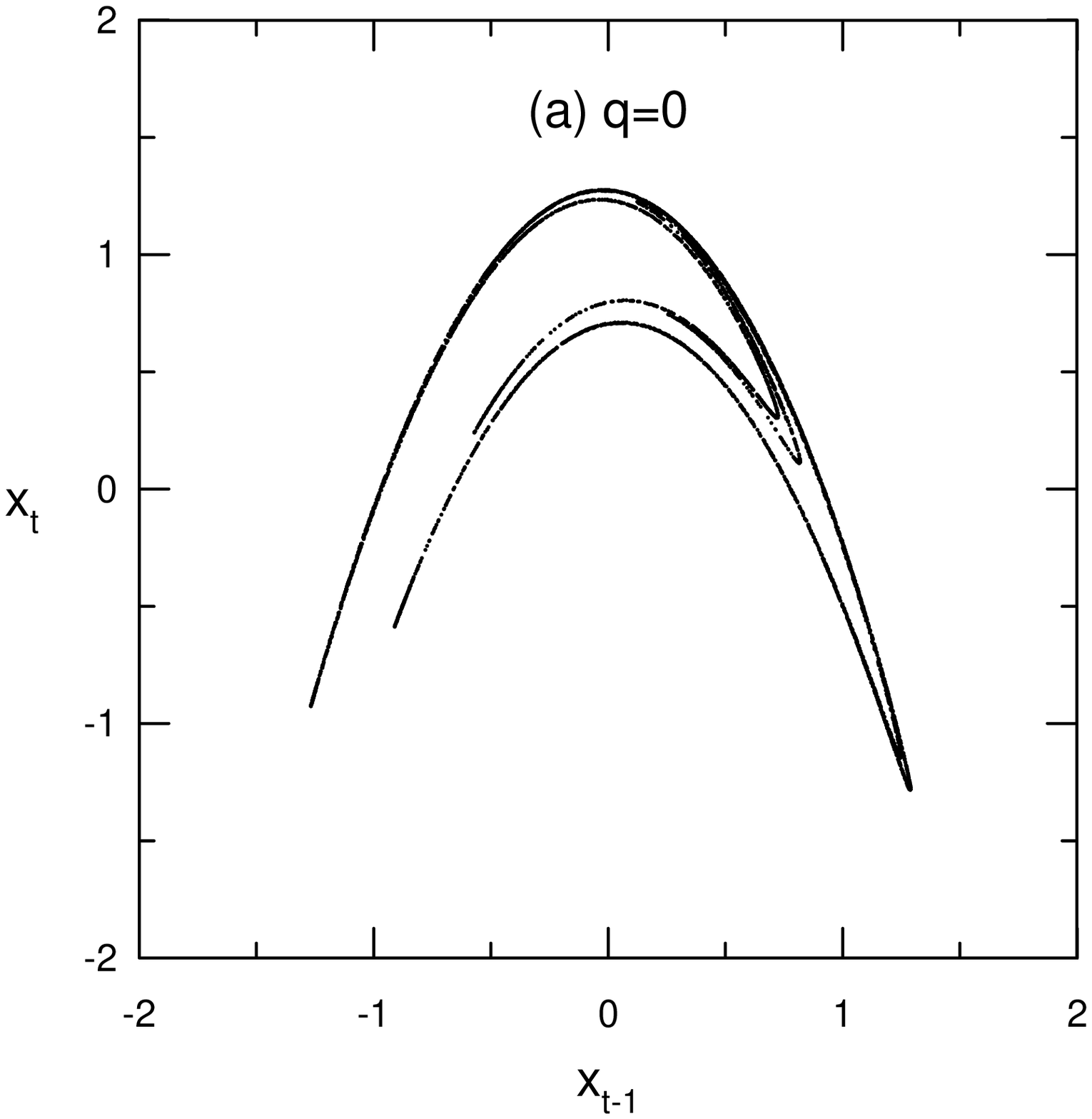}{\fullfigsize}}}
\ifcompact\vspace{2ex}\fi
\ \centerline{\hbox{\psfigure{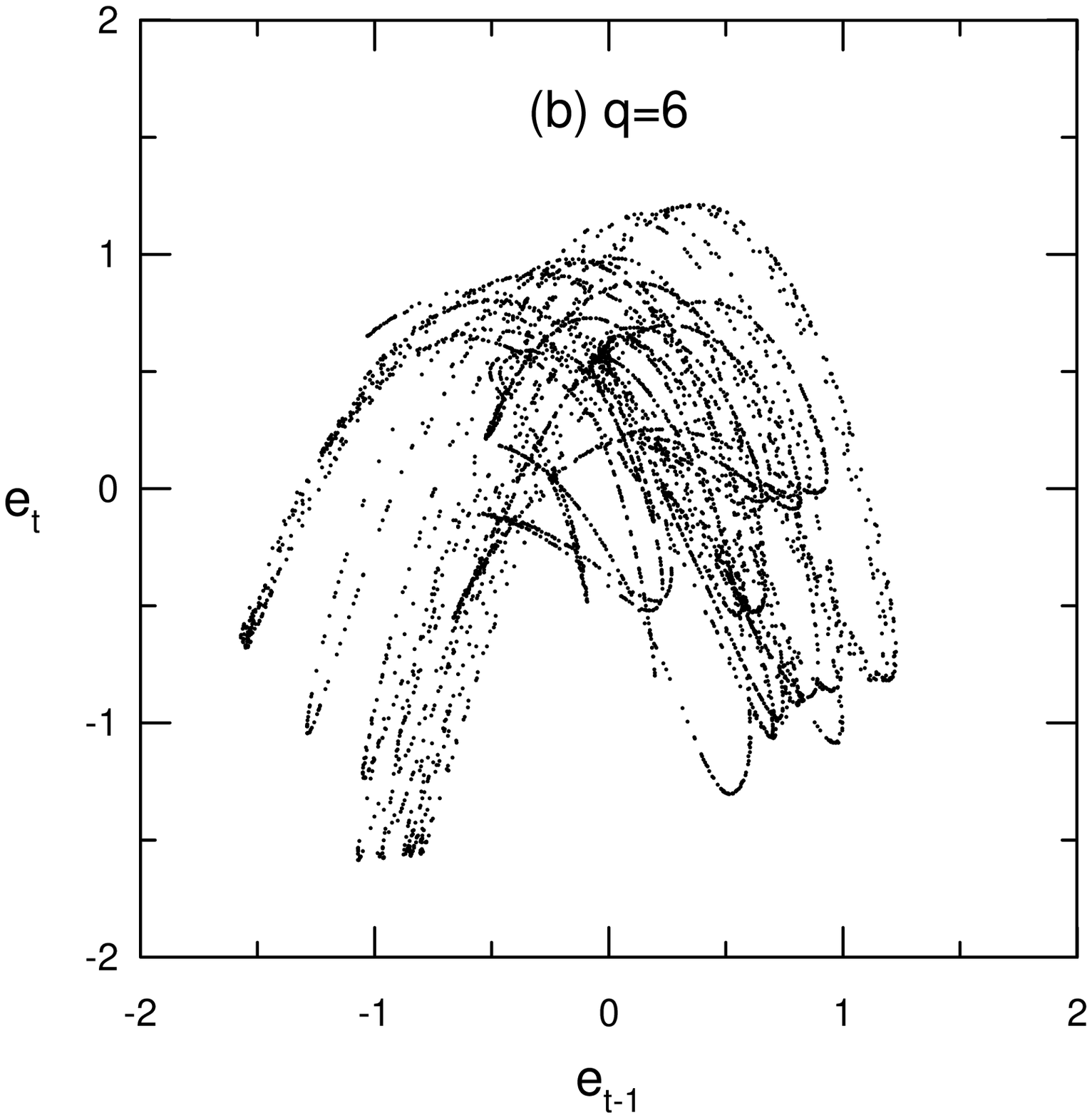}{\fullfigsize}}}
\kaption{Effect of bleaching on a time series
derived from the $x$ values of the H\'enon map.  {\bf (a)} The
unfiltered data corresponds to $q=0$.  {\bf (b)} A $q=6$ filter
distorts the attractor considerably, and hides the determinism that is
evident in the raw data.  }
\label{fig-henon-bleach}
\end{figure}

\begin{figure}[bhtp] \centerline{\hbox{
\psfigure{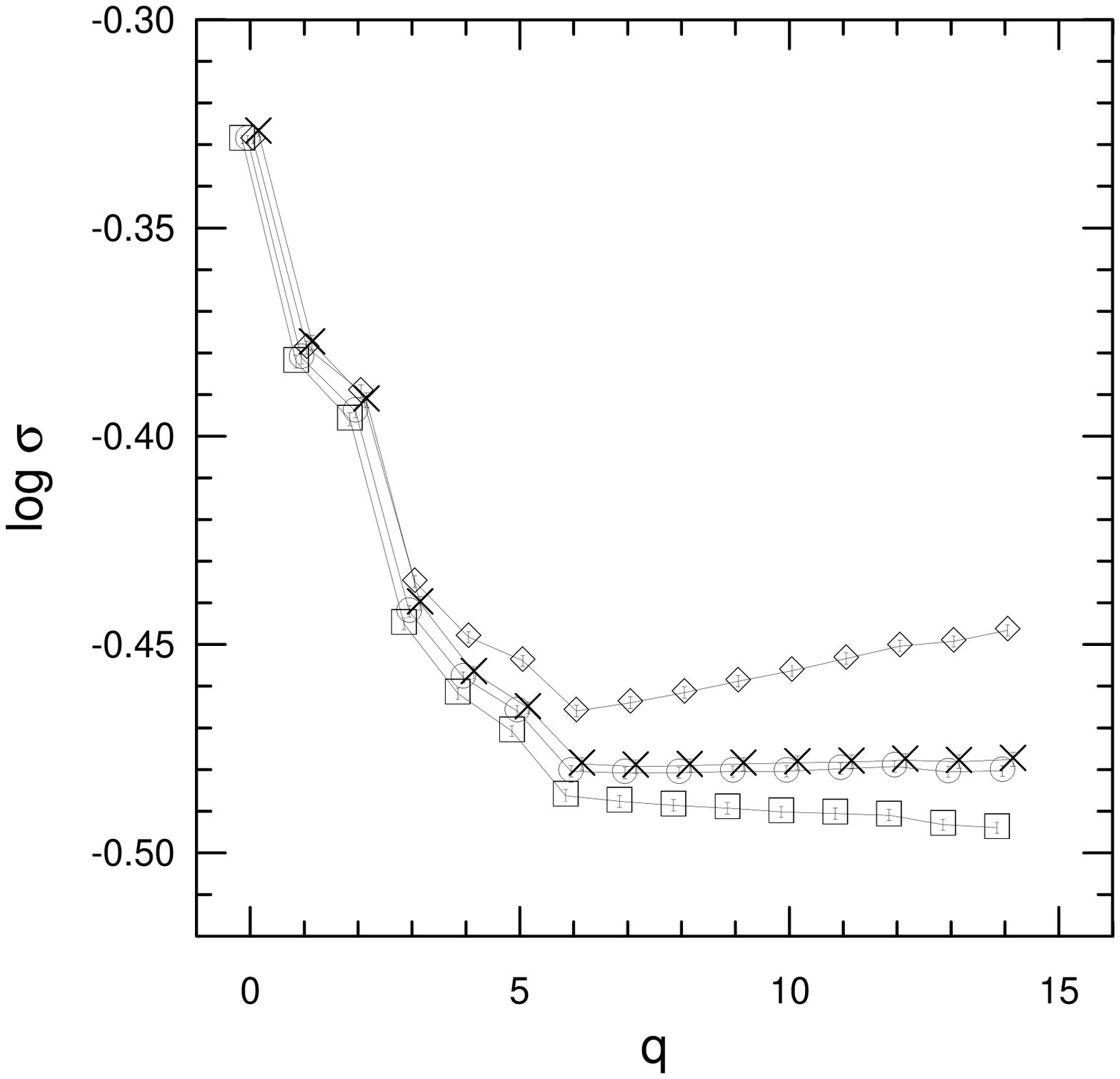}{\fullfigsize} }}
\kaption{Four measures of goodness of fit are plotted as a function
of $q$ for the H\'enon map with $N=1024$ points.  The in-sample
fitting error ($\Box$) decreases monotonically with increasing $q$
because there is no penalty for more parameters and no guard against
overfitting.  The out-of-sample fitting error ($\times$) and the
Akaike Information Criterion ($\circ$) both show a leveling-off at
$q=6$, while the Schwarz criterion ($\diamond$) indicates a definite
minimum at $q=6$, suggesting that an order 6 fit is optimal with this
many data points.  The Akaike curve is $\log\sigma + q/N$ where
$\sigma$ is the in-sample rms fitting error, and the Schwarz curve is
$\log\sigma + (q\log N)/2N$.  }
\label{fig-henon-aic}
\end{figure}

While the fit is based on the best auto-regressive (AR) {\it model},
the linear map that takes $x_t$ to $e_t$ in \eq{bleach} is a
moving-average (MA) {\it filter\/}; that is, it is a nonrecursive,
finite order, or finite-impulse-response (FIR), filter.  Strictly
speaking, it will not
change the structure of the attractor for finite
$q$~\cite{Brock86c,Sauer91,Broomhead92,Sauer93,Isabelle92}.  For
example, if $x_t$ lies on a strange attractor, then $e_t$ will lie on
an attractor of the same dimension.  This is not true of an AR filter,
which can increase the dimension of the
attractor~\cite{Badii88,Mitschke88}.  Actually,
it {\em is} possible for
a nongeneric MA filter to reduce the dimension, by ``undoing'' an AR
filter's increase~\cite{Broomhead92,Sauer93}.

However, as \fig{henon-bleach} shows, the effect of bleaching the
H\'enon attractor~\cite{Henon76} is to distort the attractor
considerably, and to make its low-dimensionality much less evident.
The order of the model, $q$, is generally chosen by some criterion
which trades off the variance of the residuals (in-sample error of
fit) against a penalty for number of parameters.  In \fig{henon-aic},
we plot Akaike's information criterion (AIC)~\cite{Akaike74},
Schwarz's criterion~\cite{Schwarz78}, and out-of-sample error
\ignore{(The equivalence of AIC and
cross-validation has been pointed out by
Stone~\protect\cite{Stone77}).} as a function of $q$, and show that
$q=6$ is a good choice for the H\'enon map with $N=1024$ points.

\ignore{
\section{Correlation dimension}
\label{sec-corr-dimension}

Although the effect of bleaching is essentially to distort the
original attractor, a particularly egregious aspect of this distortion
is to move points close together in phase space which were not
close together in the original phase space.  While this is seen
directly in \fig{de-vs-dx}, it also arises when one attempts
to estimate fractal dimension in terms of the correlation integral
of Grassberger and Procaccia~\cite{Grassberger83}.  See \fig{cr}.
What we find is that the distance scale $r_o$ below which fractal
scaling occurs is considerably reduced for the residual time series.

\begin{figure}[htbp]
\ \centerline{\hbox{
  \psfigure{cr.ps,height=\halffigsize}
  \psfigure{nu.ps,height=\halffigsize}}}
\kaption{
{\bf (a)} Correlation integral $C(r)$ for $q=0$ (solid), $q=1$ (short
dashed), $q=2$ (dash-dotted), $q=3$ (medium dashed), $q=4$ (long
dashed), and $q=5$ (dash-dotted).  {\bf (b)} Slope, $\nu(r)$, of the
correlation curves.  This slope estimates the fractal dimension.  Note
that although the bleaching does not affect the dimension as measured
by the slope at small $r$, it does affect the characteristic radius
$r_o$ below which the correct fractal scaling occurs.}
\label{fig-cr}
\end{figure}

Although we have not done the relevant numerical experiments, we suspect
that indiscriminate bleaching will have a similarly deleterious effect
on estimates of Lyapunov exponent, or upon the tests for determinism
advocated by Casdagli~\cite{Casdagli91e} and Kaplan~\cite{Kaplan92}.
}

\section{NONLINEAR MODELING}
\label{sec-modeling}

\def\e{e}
\def\pastx{\mbox{$\vec{x}_{t-1}$}}
\def\paste{\mbox{$\vec{\e}_{t-1}$}}
\def\lmap{\mbox{${\cal L}$}}
\def\nlmap{\mbox{${\cal N}$}}

\begin{figure}[bhtp]
\ \centerline{\hbox{
\psfigure{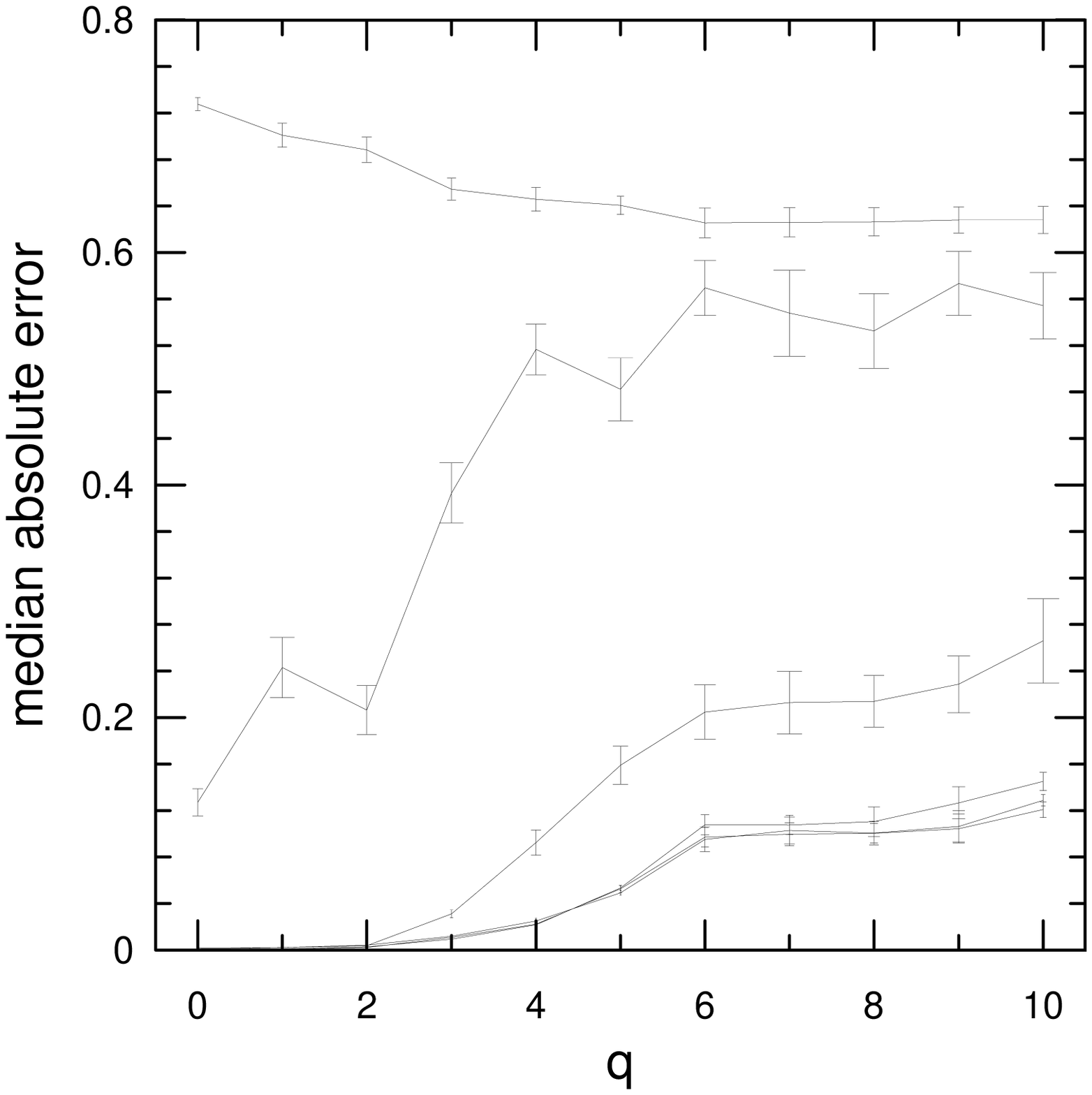}{\fullfigsize}
}}
\kaption{Modeling error of a nonlinear predictor on a time
series generated by the H\'enon map.
For $q=0$, the raw
data set is used.  For $q>0$, the $q$-th order residuals (as computed
by \protect\eq{bleach}) are used.   The top ($m=0$) curve corresponds to the
amplitude of the
order-$q$ residuals; these decrease with increasing $q$.
The curve below that is from an $m=1$ model,
and below that is $m=2$.  The curves for $m=3,4,5$ are essentially
the same.
(The error bars are based on five independent
runs with different realizations of the data.)}
\label{fig-henon-model}
\end{figure}

A very direct measure of determinism in a time series is the accuracy
of a nonlinear predictor.
We performed a numerical experiment that
involved modeling the H\'enon attractor
with a nonlinear predictor based on local-linear fits to the $k$
nearest neighbors~\cite{Farmer87}.  The results are shown in
\fig{henon-model}.  The time series contains $N=1024$
points, half of which are used for learning the nonlinear map, and the
other half for testing the goodness of the model.  We used $k=2m$,
where $m$ is the embedding dimension of the model.  In general,
increasing the embedding dimension (up to $m=3$) improves the
prediction, but increasing $q$ degrades the prediction.  Nonlinear
prediction of fully bleached data leads to errors that are in this
case two orders of magnitude larger than errors obtained by directly
fitting the raw data.

Note that for both the raw data and for the residuals, an embedding
dimension of $m=3$ is in principle adequate, since the fractal
dimension is approximately $d\approx 1.3$~\cite{Grassberger83e}, and a
theorem of Sauer\etal~\cite{Sauer91} states that as long as $m>2d$,
the embedding will almost always be sufficient.

We remark that fitting the residuals is different from a common
two-step approach to fitting data that first fits a linear model,
then fits a nonlinear model to what is left.
To make this distinction clearer, let us write
$\pastx = ( x_{t-1}, \ldots, x_{t-m} )$.
The best linear model to the time series is $\hat x_t = \lmap(\pastx)$
with $\lmap$ chosen to minimize the variance of the residuals
$\e_t = x_t - \lmap(\pastx)$.

Nonlinear modeling of the residuals
in terms of the actual past time series $\pastx$, that
is
\be
	\hat\e_t = \nlmap(\pastx),
	\label{eq-e-pastx}
\ee
permits a a full nonlinear model of the original time series:
$\hat x_t = (\lmap + \nlmap)(\pastx)$.

By contrast, nonlinear modeling of the bleached time series means
finding a nonlinear map $\nlmap'$ which estimates $\e_t$ from past
residuals $\paste$.
\be
	\hat\e_{t} = \nlmap'(\paste)
	\label{eq-e-paste}
\ee
Combining $\lmap$ and $\nlmap'$ into a full model for the original
time series is possible, but far from natural.  Furthermore, for
chaotic data, we find that the estimation errors obtained with
$\nlmap'$ are generally larger than those of the more direct $\nlmap$.

{\it Quasiperiodic data.}
	The case against bleaching depends on the time series being
chaotic.  When applied to quasiperiodic data, the ill effects of
bleaching are not evident.

\ignore{Similarly, the error of a nonlinear predictor was about the same for
raw quasiperiodic data as for bleached data; if anything, bleaching
helped.  The reason bleaching is essentially harmless in
this case is related to the ability of linear predictors to
model the time series to arbitrary accuracy if a high enough order
is chosen. (One quick way to realize this is to recall that a
quasiperiodic orbit will be almost periodic with a large period $T$,
which is a common near-multiple of the underlying periodicities,
and so a linear model $\hat x_t = x_{t-T}$ will be a very good
predictor, and of course the optimal order-$T$ linear
predictor will be even better.)
}

\begin{figure}[htbp]
\centerline{\hbox{
\psfigure{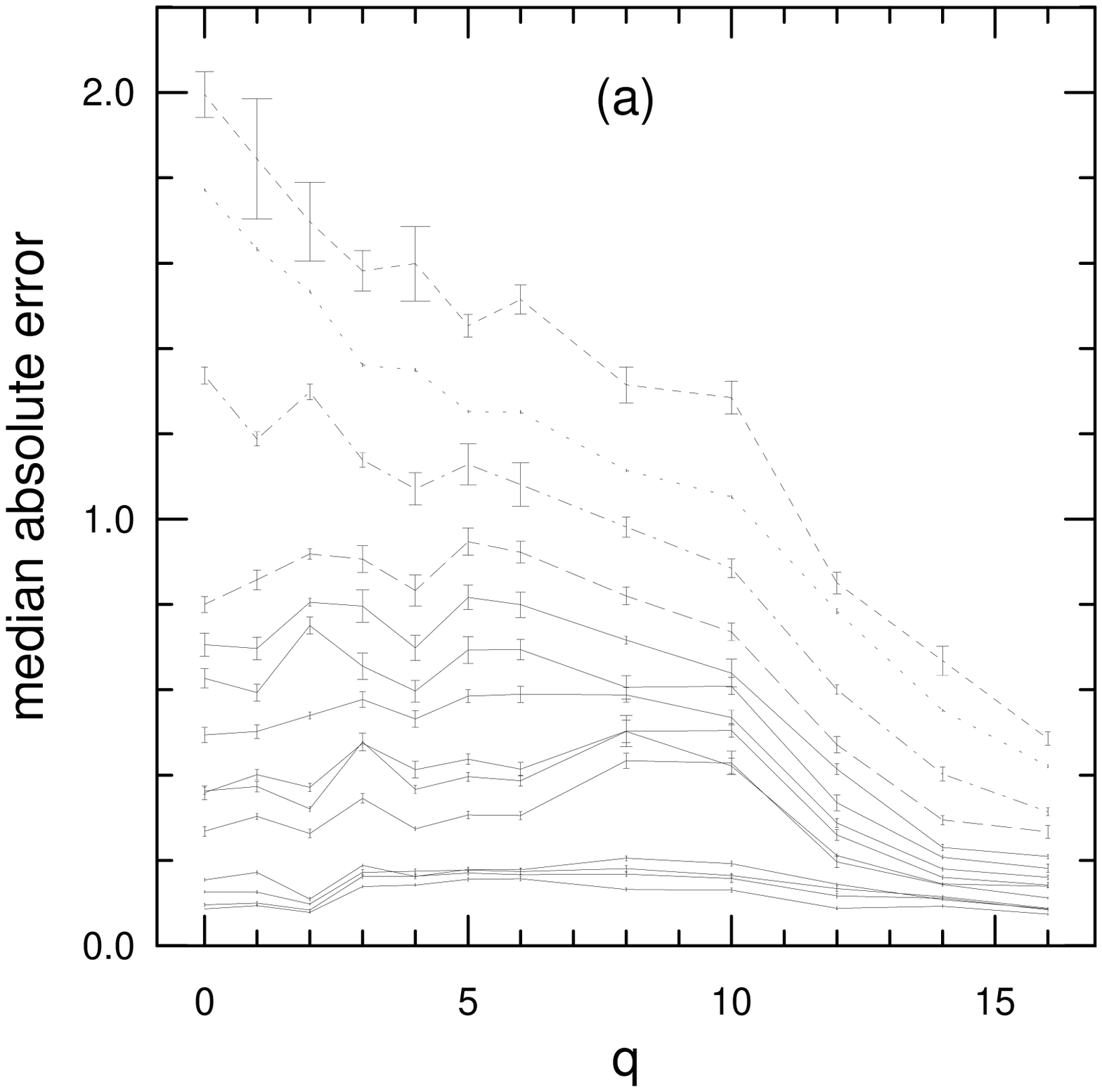}{\fullfigsize}
}}
\ifcompact\vspace{2ex}\fi
\centerline{\hbox{
\psfigure{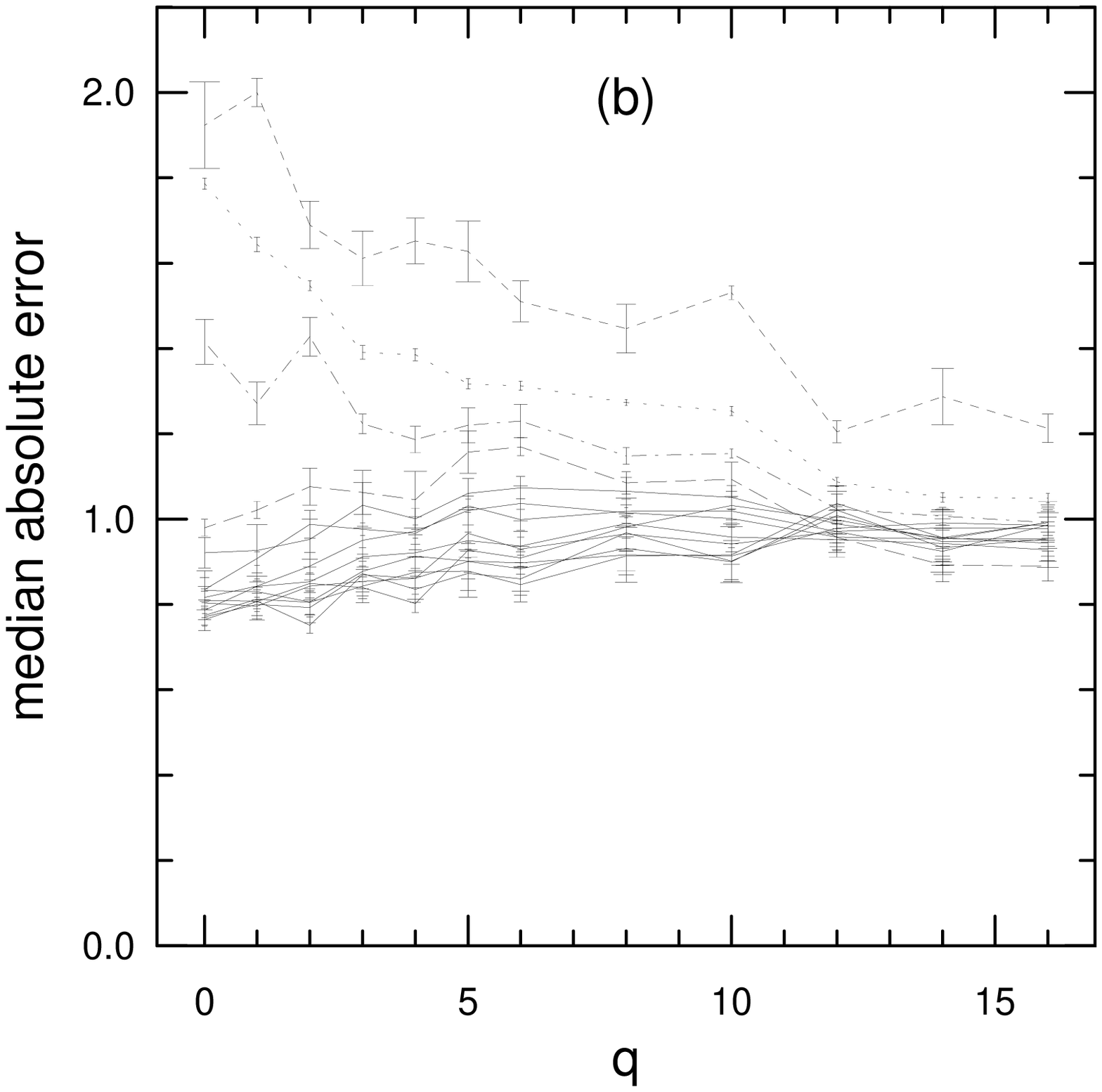}{\fullfigsize}
}}
\kaption{Similar to \protect\fig{henon-model}, but instead of using
chaotic data, we use {\bf (a)} noise-free, and {\bf (b)} noisy
quasiperiodic data.
Note that in contrast to the case of chaotic data,
the effect of bleaching is not to degrade the accuracy of a nonlinear
model, but on the
contrary to improve it.  Embedding dimensions shown are
$m=0$ (dotted line); $m=1$ (short dashed line); $m=2$ (dashed-dotted line);
$m=3$ (long dashed line); and $m=4,5,6,8,10,12,14,16,18,20$ (solid lines).
In general, the larger the $m$, the smaller the error (except the $m=0$
curve, which is actually more accurate than the $m=1$ curve).
}
\label{fig-model-quasi}
\end{figure}

For our numerical experiment, we deliberately chose an example that was
more complicated than the sum of two sine waves.  The
quasiperiodic data were generated by
a nonlinear two-frequency model with observational and dynamical
noise:
\be
	x_t = \frac{X_{1,t}+X_{2,t}+1}{X_{1,t}^2+X_{2,t}^2+0.2}
\ee
where $X_{i,t}=\sin(\phi_i + (2\pi/5)\gamma_i t + \eta_{i,t}) +
0.1\varepsilon_{i,t}$.  Here, the two mutually incommensurate
frequencies are $\gamma_1=(\sqrt{5}-1)/2$ and $\gamma_2=\sqrt{3}-1$.
Observational noise is modeled with $\varepsilon$, a Gaussian white
noise process with unit variance; and dynamical noise is
modeled as a random-walk phase drift: $\eta_t = \eta_{t-1} +
0.1\epsilon_t$, where $\epsilon_t$ is again Gaussian white noise with
unit variance.  Finally, $\phi$ is a randomly chosen initial phase.
Five time series were generated, using different starting phases
$\phi$ and a transient time $N_{\rm transient}=512$.  The time series
themselves were of length $N=512$.  Each was modeled by a local linear
map with various embedding dimensions $m$ and bleaching parameters
$q$.  We used the $k=2m$ nearest neighbors from the first half of the
data set for one-step-ahead predictions on the second half of the data
set, and computed the median absolute error.  As seen in
\fig{model-quasi}, unlike the case with chaotic data, bleaching does
not have such a debilitating effect on the modeling.

\section{DETECTING NONLINEARITY}
\label{sec-test-nonlin}

In this section, we will describe how bleaching influences statistical
tests for nonlinearity.  The motivation behind a test for nonlinearity
is sometimes simply to determine whether a linear model will capture
all of the structure in the time series.  Often, however, there is a
hidden agenda.  One may seek to detect nonlinearity as a first step in
what is ultimately a search for chaos.  Nonlinearity is certainly a
pre-requisite for chaos, but it is not the most straightforward way to
test for chaos.  A more direct approach might be to estimate the
largest Lyapunov exponent.  A positive Lyapunov exponent implies
chaos, so a positive estimate would be taken as direct evidence in
favor of chaos.  The main problem with this approach is that the
estimation of Lyapunov exponent is a nontrivial
procedure~\cite{Abarbanel91,Nychka92}, and it is difficult to quantify
the reliability of the estimate.  Testing only for nonlinearity may
not be as direct a test for chaos (the disadvantage being that a positive
identification of nonlinearity does not imply chaos), but it can be
done far more reliably than trying to compute a Lyapunov exponent.

\ignore{
In this section, we will describe how bleaching influences statistical
tests for nonlinearity.  Since nonlinearity is a prerequisite for
chaos, this is often a first step in what is ultimately an attempt to
detect chaos itself.  Arguably the most straightforward test for chaos
is the estimation of largest Lyapunov exponent from the time series.
Since a positive Lyapunov exponent implies chaos, estimating a
positive value is taken as evidence that chaos is present.  The main
problem with this approach is that the estimation of Lyapunov exponent
is a nontrivial procedure~\cite{Abarbanel91,Nychka92}, and it is
difficult to quantify the reliability of the estimate.  By contrast,
testing for nonlinearity can be done much more reliably than trying to
compute a Lyapunov exponent.  The disadvantage here is that a positive
identification of nonlinearity does not necessarily imply chaos.
}

Bleaching provides a conceptually simple approach
to testing for nonlinearity in time series.  Since the residuals of a
bleached time series have no linear correlations, any correlations
that are found in the residuals must be nonlinear.  In particular,
testing the residuals against IID (independent and identically distributed)
is equivalent to testing the original
time series for nonlinearity.

This is the basis of the Brock--Dechert--Sheinkman (BDS)
test~\cite{Brock86} (see Ref.~\cite{Brock91d} for a recent and more
complete exposition), the tests for chaos described by
Hsieh~\cite{Hsieh91} (though with the financial time series of
interest here, there is little autocorrelation to begin
with~\cite{Hsieh89}), a neural-net-based test for ``neglected
nonlinearity''~\cite{LeeTH92}, as well as a variety of classical
nonlinearity tests~\cite{McLeod83,Keenan85}, many of which are
reviewed in Tong~\cite{Tong90}.  To be fair, not all of these tests
were designed with the idea of looking for chaos.  Our point is that
those tests which have bleaching as their first step will have low
power when the test data is chaotic.  We should also be careful to
note that the test proposed by Tsay~\cite{Tsay86}, though it involves
residuals, also makes use of the original data.  It has the flavor of
\eq{e-pastx} as opposed to \eq{e-paste}, and unlike purely
residual-based statistics, it may not suffer the same loss of power
against chaotic time series.

Instead of comparing residuals to IID, a more direct approach is to
compare the original data to surrogate data sets which mimic the
linear correlations in the original time series, but which are
otherwise
random~\cite{Theiler91b-x,Kaplan90,Tsay92,Smith92,Theiler91c}; in the
statistical literature, the approach is often identified as a
bootstrap.  There is some discussion of the
connection between the surrogate data approach and the classic
bootstrap in Refs.~\cite{Theiler91c,Theiler93b-x}; the interested
reader should also consult Refs.~\cite{Tsay92,Efron86} for pointers
into the relevant literature.

A discriminating statistic (which for chaotic
processes is often chosen to be a dimension or Lyapunov exponent
estimator, or the error in a nonlinear predictor, but in general can be any
function that maps a full time series into a single number) is computed for
each of the surrogates and for the original data set.  If the number
obtained for the original data set is significantly different from
those obtained for the surrogate data sets, then a null hypothesis of
linearly correlated noise can be rejected.  A crude (and cheaply computed)
measure of how significantly different the original is from the surrogate
data is given by the number of ``sigmas'':
\def\stat{Q}
\be
	{\rm sigmas} = \frac{|\overline{\stat}_{\rm surrogate}
	-\stat_{\rm original}|}
    {\sigma_{\rm surrogate}}
\ee
Here $\stat_{\rm original}$ is the value of the discriminating
statistic for the original data set, and $\overline{\stat}_{\rm
surrogate}$ and $\sigma_{\rm surrogate}$ are the mean and standard
deviation, respectively, of the discriminating statistics computed for
the surrogate data sets.  We remark that this is a heuristic measure.
Properly, one should compute the probability (also called the
$p$-value) of mis-identifying a linear time series as nonlinear.  One
way to estimate $p$ is from the percentile ranking of $\stat_{\rm
original}$ in a sorted list of all the $\stat$ values.  Only when the
$\stat$ statistic has a distribution of some previously assumed form
(usually Gaussian) can the $p$-value can be computed directly from the
number of sigmas.  In general, though, the more sigmas, the smaller
the $p$-value, and the more powerful the statistic.  We will be
using sigmas as an inexpensive measure of relative power.

Formally, the method of surrogate data provides a measure of
statistical confidence that the null hypothesis is false; informally,
it can be used as a control experiment to assess whether the
measurement of a given nonlinear property is being fooled by simple
linear correlation in the time series.

Our approach will be to compare the power of different tests for
nonlinearity when the form of that nonlinearity is chaos.  In
statistical terminology, the null hypothesis is linearly correlated
noise, and the alternative hypothesis is chaos.  If the alternative
hypothesis is a specific chaotic process, one can imagine designing
very sensitive tests for distinguishing this process from the null.
For the broad class of chaotic processes (and especially for the even
broader class of nonlinear processes that may or may not be chaotic),
the notion of an optimal design ceases to be well-posed.  The emphasis
here, however, will not be on finding the most powerful tests for
nonlinearity; instead we will concentrate on the simpler question of
how bleaching affects the power of existing tests when the alternative
is chaos.

\begin{figure}[bhtp]
\centerline{\hbox{
\psfigure{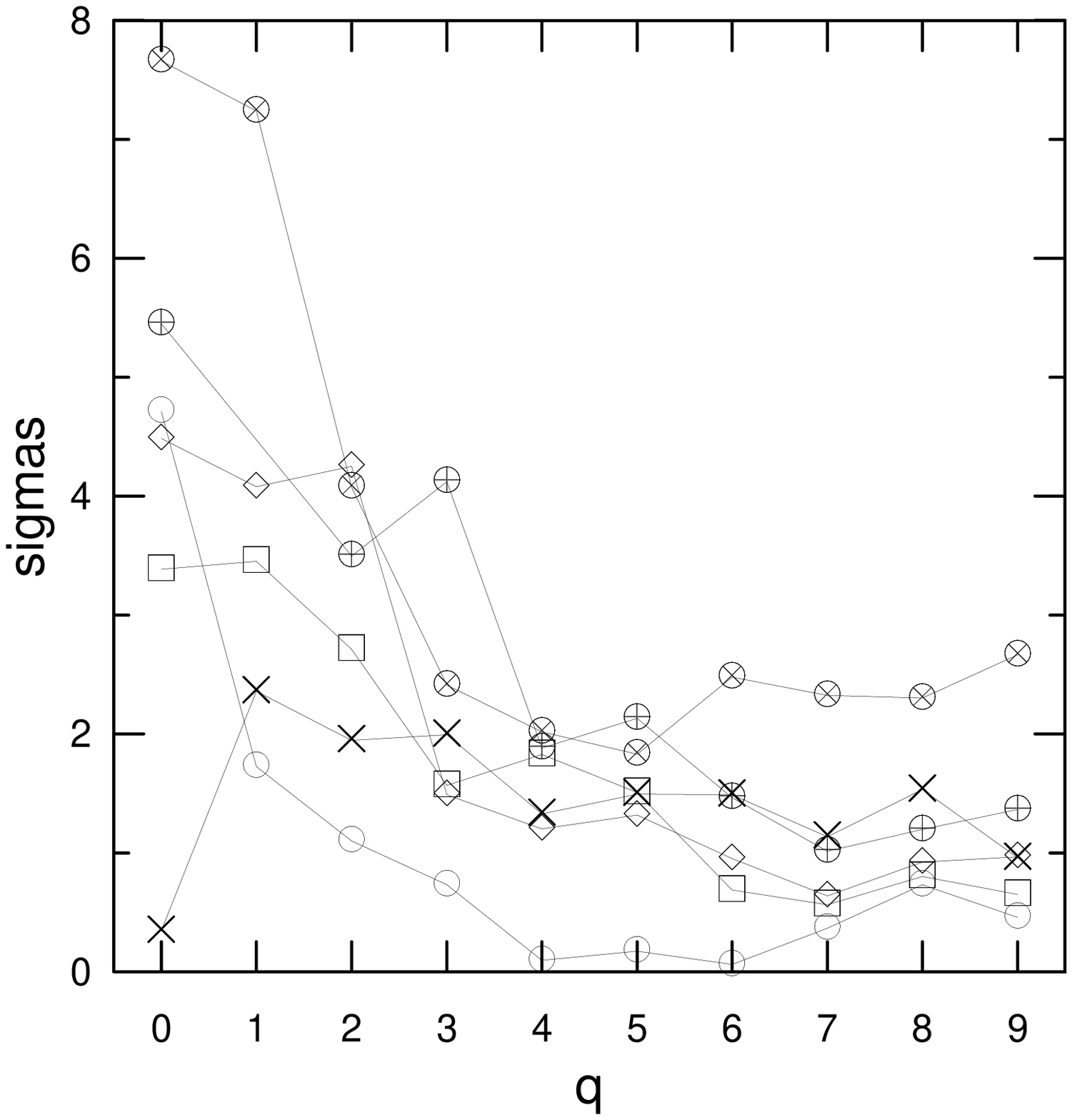}{\fullfigsize}
}}
\kaption{Significance of rejection of a null hypothesis of linearly
correlated noise versus bleaching parameter $q$ for a variety
of discriminating statistics:
modified BDS ($\oplus$),
a simple skew statistic ($\otimes$),
estimated correlation dimension ($\circ$),
the correlation integral itself ($\Box$),
local linear forecasting error ($\diamond$),
and modified McLeod-Li ($\times$).
For the experiment in this figure, we used a
time series of $N=1024$ points
obtained by summing four independent realizations of time series
from the H\'enon
map.  The dimension, BDS, and forecasting statistic used an embedding
dimension of $m=3$.   Although this is clearly too small to see the
full dynamics in the time series, for the purpose of finding evidence
for nonlinearity from a series of this length, the value $m=3$ was
empirically found to give the most significance (the skew and modified
McLeod-Li statistics do not require an embedding).
All of the discriminating statistics (except the modified
McLeod-Li)
show evidence of nonlinearity at the three sigma level for unbleached
data ($q=0$), and all of them {\em fail} to show evidence of nonlinearity
at the three sigma level for the fully bleached data ($q\ge6$).}
\label{fig-significance}
\end{figure}

In the numerical experiment shown in \fig{significance}, significance
was computed for a variety of discriminating statistics on a chaotic
time series and on time series obtained by bleaching with ever larger
values of $q$.  By and large, the significance was found to decrease
with increasing $q$.  We also
performed some experiments with quasiperiodic data (not shown), and we
found that bleaching did not noticeably alter the ability of the
surrogate data method to detect nonlinearity.  We remark, however,
that attempting to distinguish nonlinearity in quasiperiodic data is a
very fussy issue.  Stable limit cycles and limit tori arise only in
nonlinear systems, yet the absence of chaos implies that linear models
(of sufficiently high order) can in principle do as well as nonlinear
models.  This issue is discussed in further detail in
Ref.~\cite{Theiler93b-x}.

\ignore{\subsubsection*{Skew statistic}}
In the method of surrogate data, just about any nonlinear statistic
can be used.  For example, we have found a very simple measure of
nonlinearity that is motivated by the fact that linear time series
have symmetric rise and fall times; the asymmetry in the derivative
can be measured by a simple skew statistic, $\langle (x_t-x_{t-1})^3
\rangle$.  For the experiment in \fig{significance}, it is this
statistic which we found most sensitive to the nonlinear structure in
the time series.  (See Tsay~\cite{Tsay92} for further discussion of
this statistic.)\comment{There is also a Ramsay-Rothman statistic that
is relevant here.}

\subsection{COMPARISON TO BDS}

Brock, Dechert, and Scheinkman~\cite{Brock86} developed a statistic to
test for nonlinearity based on the correlation integral of Grassberger
and Procaccia~\cite{Grassberger83}.  This is, to our knowledge, the
first statistically rigorous test to exploit the ``new paradigm'' of
deterministic chaos as an alternative hypothesis.  To define the BDS
statistic for a time series of $N$ points, first define $C_m(N,r)$ as
the m-dimensional correlation integral
\be
	C_m(N,r) = \frac{2}{N(N-1)}\sum_{i=1}^N\sum_{j=1}^{i-1}
	\prod_{k=0}^{m-1} \Theta(r - |x_{i+k} - x_{j+k}|)
\ee
where $\Theta(x)$ is the Heaviside step function; it is one for
positive $x$, and zero otherwise.  For IID data, in the limit $N\to\infty$,
one expects
$C_m(N,r) \approx C_1(N,r)^m$.  In particular, the BDS statistic
\be
	Q_{\rm BDS} = \sqrt{N}\left[C_m(N,r)-C_1(N,r)^m\right]
\ee
will for IID data converge to a normal distribution with zero
mean and fixed variance.  The variance can be estimated from the data, but
for our purposes, we find it convenient to estimate the variance
using Monte-Carlo simulation.

\begin{figure}[bhtp]
\centerline{\hbox{
\psfigure{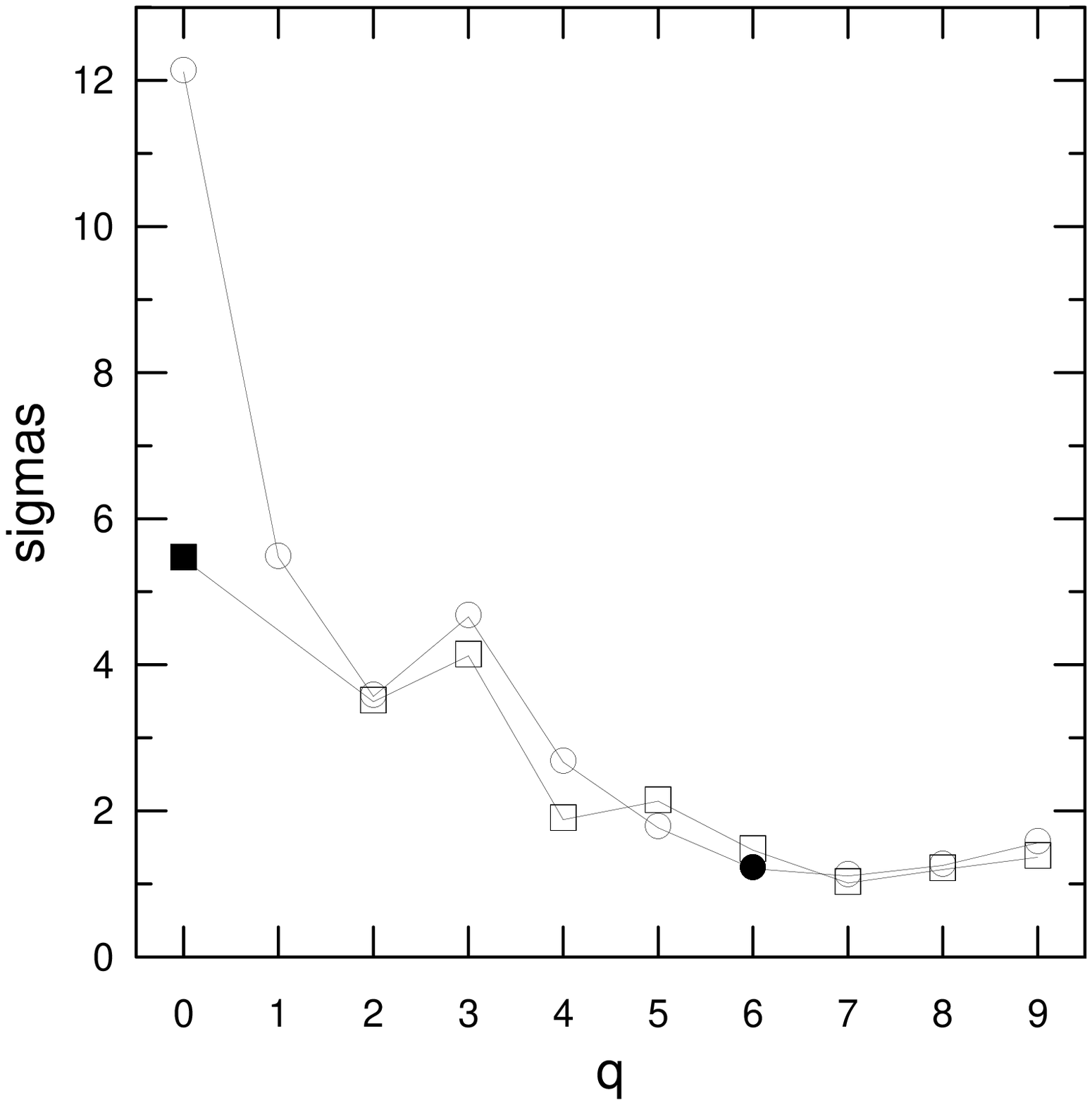}{\fullfigsize}
}}
\kaption{Significance of the BDS statistic as a function of
bleaching $q$.  The time series is $N=1024$ points obtained
by adding four independent realizations of H\'enon time
series.  As with the single H\'enon time series, full bleaching
occurs at $q=6$. The circle ($\circ$) curve uses BDS
to test against a null of IID noise.  Not surprisingly, the null
is easily rejected for unbleached data, because there are both
linear and nonlinear correlations, and the test doesn't distinguish
them.  However, applying the test to bleached data, we
find little evidence to reject the null of IID residuals.
On the other hand, the square ($\Box$) curve uses the BDS statistic
as part of a surrogate data algorithm to test directly against
the null hypothesis of linearly correlated noise.  This test is less
significant at $q=0$, but that's because it is testing against a more general
null.  It too loses significance as $q$ increases.  But what should be
compared here is the $q=0$ square (\protect{\rule{1ex}{1ex}}) point,
and the $q=6$ circle
($\bullet$) point; the former is significant at the five sigma level,
while the latter is not significant.  The former uses the BDS to test the
raw time series against a null of linearly correlated noise; the latter
uses BDS (as it was originally intended) to test residuals against IID;
though the two tests are formally equivalent, the direct test that
avoids bleaching is the more powerful.}
\label{fig-bds}
\end{figure}

In particular, we use $Q_{\rm BDS}$ as the discriminating statistic in
the scheme of surrogate data. We find that as a discriminating
statistic, it is quite powerful.  However, when it is applied to
bleached data, it loses its original power.  We suggest therefore that
the BDS statistic should not be applied to residuals and compared
against IID, but instead should be applied to the original data and
compared against the appropriate surrogates (see \fig{bds}).

\ignore{We should emphasize that we are applying this statistic to time
series that are known to be low-dimensionally chaotic.  The statistic
was originally developed in the context of financial time series.
There were historical and pedagogical, as well as scientific reasons
for this approach;  indeed, Brock was aware that residuals could
hide chaos in his 1986 paper.}

\subsection{COMPARISON TO MCLEOD-LI}

One of the most straightforward conventional approaches to testing for
linearity in a time series is to look at the autocorrelation
of the squared residuals.  If the residuals truly are IID, then their
squares will be IID, and therefore, the squares will have zero
autocorrelation.  In particular, the statistic based on sample
autocorrelation of the squares
\be
	Q_{\rm ML} = N(N+2)\sum_{k=1}^m \frac{1}{N-k}r_k^2,
\ee
where
\be
    r_k = \frac{\langle e_t^2e_{t-k}^2 \rangle - \langle e_t^2 \rangle^2}
         {\langle e_t^4 \rangle - \langle e_t^2 \rangle^2}
\ee
is the autocorrelation of the squared time series,
will for IID data converge as $N\to\infty$ to a well-defined
distribution.
This is a particular case of the McLeod-Li~\cite{McLeod83}
statistic~\cite{choice-of-mcleod-li}.
As in the case of the BDS statistic,
we can apply the statistic to unbleached data by simply
using $x_t - \langle x_t \rangle$ in place of $e_t$ in the above
formula.  However, at least for the numerical experiment in
\fig{significance}, we found that this statistic was the weakest of
our tests for nonlinearity.

\ignore{\subsubsection*{Modified McLeod-Li}}

\begin{figure}[bhtp]
\ \centerline{\hbox{
\psfigure{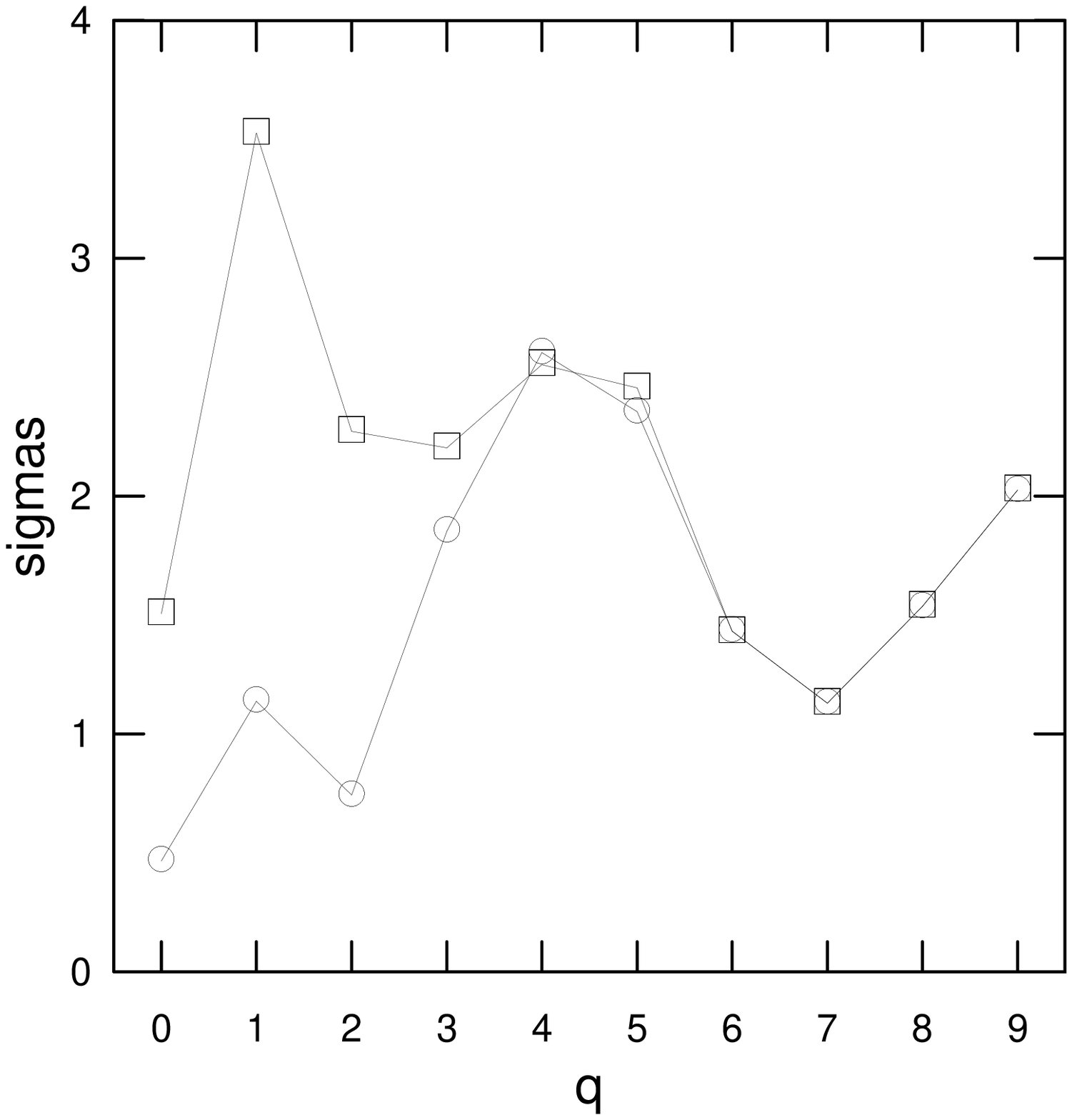}{\fullfigsize}
}}
\kaption{Significance of nonlinearity for a time series of
$N=1024$ points
obtained by summing
four realizations of the H\'enon map, using the McLeod-Li
statistic ($\circ$) and a modified version of McLeod-Li ($\Box$)
described in \protect\eq{improved-mcleod-li}.
}
\label{fig-improved-mcleod-li}
\end{figure}

Further, the McLeod-Li statistic seems to
{\em improve} when the data set is bleached. This can be understood
intuitively by realizing that the autocorrelation of the squared
time series involves very large values (and therefore, very large
variances).  One natural way to reduce these values is with
the following modification:
\be
	Q_{\rm MML} = N(N+2)\sum_{k=1}^m \frac{1}{N-k}(r_k-A_k^2)^2
	\label{eq-improved-mcleod-li}
\ee
where $r_k$ is the autocorrelation in the squared time series
(as before), and $A_k$ is the autocorrelation of the original
time series.
\be
	A_k =
    \frac{\langle x_tx_{t-k} \rangle - \langle x_t \rangle^2}
    {\langle x_t^2 \rangle - \langle x_t \rangle^2}.
	\label{eq-autocorr}
\ee
The idea is
to ``subtract off'' that much of the autocorrelation of the squares
which can be attributed to the autocorrelation in the original
time series.  \fig{improved-mcleod-li} shows that the new statistic
is more powerful when used with surrogate data; and for the data
set under consideration, is optimal for a small value of $q$.

\section{SOME GENERAL REMARKS ON LINEAR FILTERING}
\label{sec-misc}

	In the case of the H\'enon attractor, bleaching is found to be
detrimental both to nonlinear modeling and to detecting nonlinearity.
But it would be incorrect to assume that all linear filtering is in
all cases bad.  Given a particular data set, and a particular
nonlinear task, one expects that there is a particular linear
prefiltering that will optimize the performance at the given task.
The theme of this \whatthisis\ is that the particular linear filter
that corresponds to bleaching is rarely optimal, and usually makes
things worse.

	In this section, we will give two examples of situations that
arise frequently in practice.  In both cases, linear prefiltering
is seen to be advantageous, but in neither case is {\em full} bleaching
recommended.

\subsection{UNFILTERING FILTERED DATA}
\label{sec-unfilter}

	A natural example is to begin with a known chaotic time series,
and then to low-pass filter the data, so as to introduce
a lot of linear correlation in the data.  For example, if
$h_t$ is a chaotic time series, and $|\alpha|<1$,
then
\be
	x_{t} = \alpha x_{t-1} + h_t
	\label{eq-hen-ar}
\ee
gives a time series $x_t$ which for $\alpha$ near 1 is dominated by
the linear component~\cite{thank-brock}.
While this example may appear at first sight contrived, it
represents a very common physical occurrence: the observation of
a natural phenomenon through a low-pass filter.  For instance, a
resistance $R$ and capacitance $C$ between the probe and
the phenomenon being measured leads to a characteristic time of $RC$,
and corresponds to $\alpha=e^{-1/RC}$ in \eq{hen-ar}.  This is
certainly the situation for the example of scalp-based
\comment{as opposed to direct invasive probes drilled through
the skull and stuck right into the brain --ugh!} measurements
of brain electrical activity, as in the electroencephalogram (EEG).

\begin{figure}[htbp]
\centerline{\hbox{
\psfigure{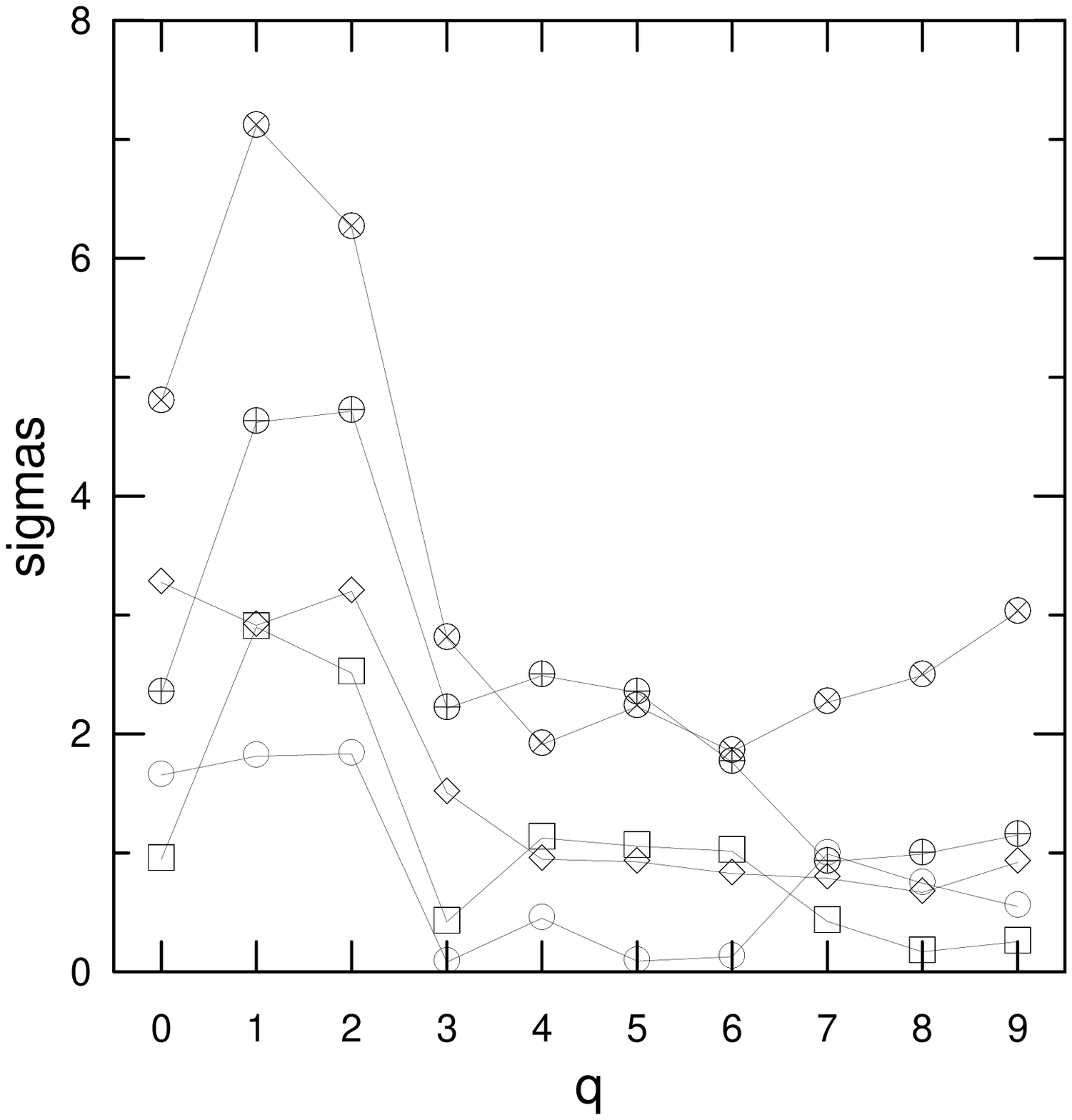}{\fullfigsize}
}}
\kaption{Here, the chaotic time series is obtained by AR filtering
($a=0.9$) a time series of four independent H\'enon maps summed
together.  The time series is bleached
at several values of $q$. As before, the embedding dimension
is $m=3$. Some bleaching (at $q=1$ or $2$) leads
to significant evidence for nonlinearity, but full bleaching
(at $q=7$ for this time series) gives time series with no detectible
nonlinear structure.  Here, the
discriminating statistics used were: modified BDS ($\oplus$),
skew ($\otimes$), correlation dimension ($\circ$), correlation
integral ($\Box$), and local linear forecasting error ($\diamond$).}
\label{fig-hen-ar}
\end{figure}

	In this case, it can be advantageous to digitally filter the
observed time series to counteract the effect of the filter through
which the data were observed.  However, it is {\it still} not
recommended to fully bleach the data!  In particular, \fig{hen-ar} shows
that for a sum of four H\'enon time series,
prefiltered with $\alpha=0.9$, the
optimum amount of bleaching is given at $q=1$ or $2$.  However, from
the point of view of linear modeling, $q=7$ is the
``proper'' amount of bleaching
for this time series (based, as in \fig{henon-aic}, on AIC, Schwartz,
and out-of-sample error criteria).  At $q=7$, the significance of the evidence
for nonlinearity is negligible.  The evidence at $q=0$ is not very
significant (depending on the discriminating statistic),
so there is a real advantage to a ``little'' bleaching to remove a dominating
linear component.

	In a more practical situation, if one is seeking evidence for nonlinearity
in a time series of sea levels, it can be advantageous to ``filter out''
the daily and monthly tides which dominate the variations~\cite{Berge90}.
We also note that Townshend~\cite{Townshend91-x} reported improved modeling
of speech signals after linear filtering; we suspect that this is due
to the dominant underlying periodicity of these signals. \ignore{In any
case, it is not clear that single spoken sentence can be considered a
stationary time series, and it certainly is not clear that the time
series is chaotic!}

\subsection{BLEACHING OVERSAMPLED DATA}

	Data which are sampled at a much higher rate than that of the
underlying physical process will have very little power in the
high frequencies.  Since the effect of bleaching is to achieve
equal power at all frequencies, the effect on oversampled data
is to grossly amplify the high frequency behavior.

\begin{figure}[htbp]
\centerline{\hbox{
\psfigure{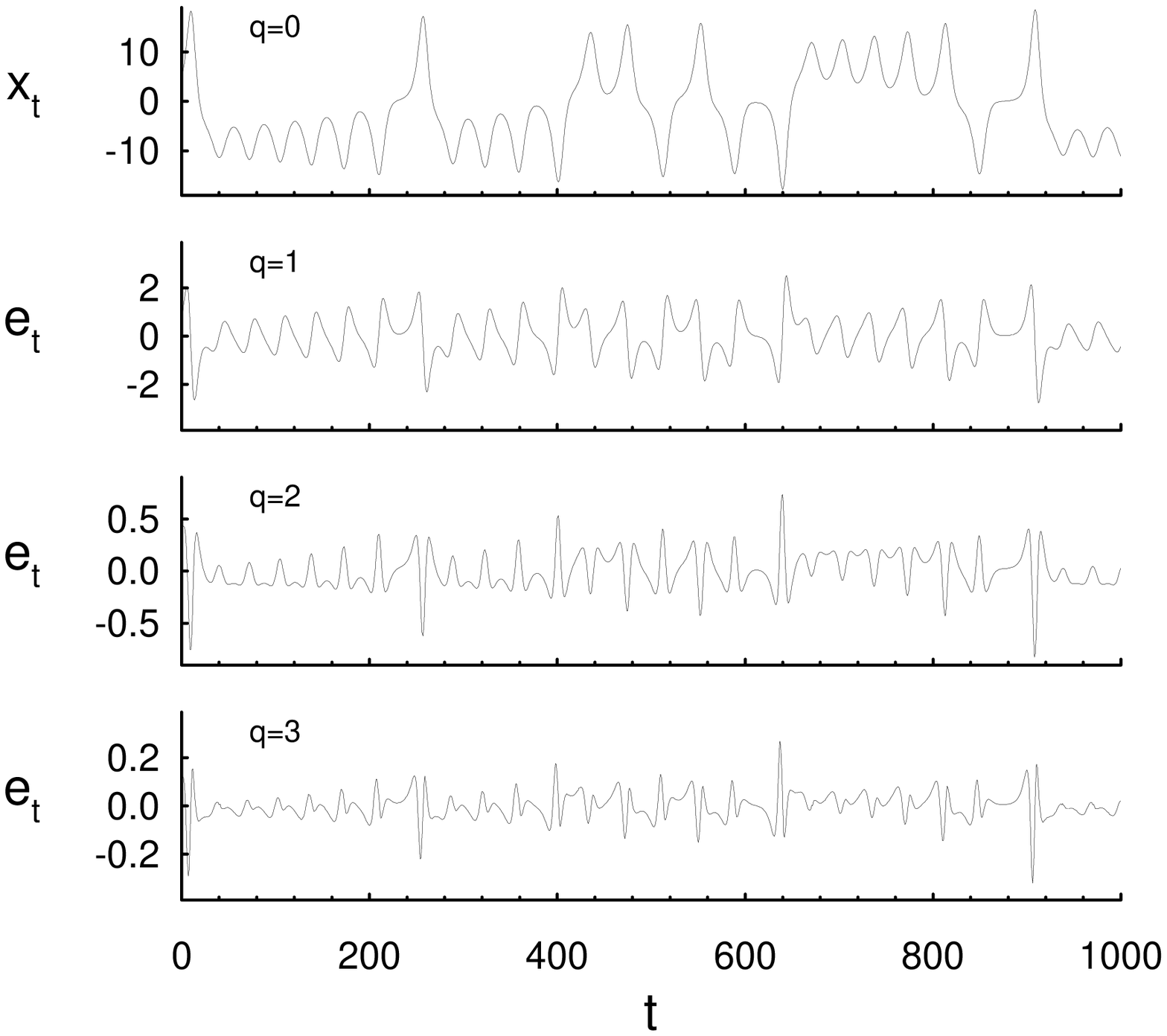}{\fullfigsize}
}}
\kaption{Bleaching oversampled data; the Lorenz time series was
sampled at a rate $\Delta t=0.02$, and residuals were computed for
$q=1,2,3$.  While bleaching significantly reduced the magnitude of
the residuals, it produced in its place a very ``spikey'' time series
that is more more difficult to
analyze than the raw data.
}
\label{fig-lz-bleach}
\end{figure}

For noise-free oversampled data, the residuals will have very small
amplitude compared to the original data, but the enhancement of the
high frequencies will lead to
very irregular and
``spikey'' dynamics.  \fig{lz-bleach} shows this effect with the Lorenz
attractor~\cite{Lorenz63}; data from the Rossler
attractor~\cite{Rossler76},
which has a more pronounced periodicity, shows the effect even more
severely.

For a time series which is oversampled from a continuous flow but
whose measurement is contaminated with uncorrelated additive noise,
the effect of bleaching is to amplify the noise.
Because this situation is so
common in physics experiments, it is sometimes difficult for
physicists to imagine why one would ever want to bleach data in the
first place.  The physicist's intuition in this case is absolutely
correct.  This is an example where it is not only unwise to bleach the
data, but it is often helpful to filter the data with a low-pass
filter, making it {\it less white} than the original signal.

	\ignore{Now, how did we know that the high frequency stuff was noise,
and could therefore just be filtered out?  Did we try and model the high
frequency stuff with low-dimensional dynamics unsuccessfully first, or did
we use some knowledge of the physics.  Suppose instead of adding white
noise, a tent-map dynamics was added; what do we do then?  How much of
this can be recast in the language of SVD style embedding; or perhaps
canonical variate analysis is appropriate, since both input and output
are so noisy?}

\section{THEORETICAL CONSIDERATIONS}
\label{sec-theory}

We have emphasized numerical experiments in this exposition, partly
because these provide graphic demonstrations of the phenomena, but
also because we do not have a good ``theory'' of why bleaching should
be so detrimental to so many different aspects of nonlinear time
series analysis.  Intuitively, the linear filtering replaces the
current state with a linear combination of states at previous times,
and the effect of this combination is to confuse the meaning of the
current state; this intuition is  made more precise in the following
section.

\subsection{LIMIT OF INFINITE DATA}

	When an infinite amount of data is available, then conditions such as
those plotted in \fig{henon-aic} do not put a cap on the order of the
bleaching filter.  That is, one may
may have $q\to\infty$ in \eq{bleach}.  This is no longer a
finite-impulse-response (FIR) filter, but is an
infinite-impulse-response (IIR) filter, and so the theorems of
Refs.~\cite{Brock86c,Sauer91,Broomhead92,Sauer93,Isabelle92} no longer
apply.  The filtered time series is no longer guaranteed to preserve
the nonlinear invariants, such as attractor dimension, of the original
time series.  In this section, we describe conditions under which a
particular invariant, the Lyapunov dimension, is altered.  We speculate
that these conditions will apply to more general invariants as well.

The Lyapunov dimension was defined by Kaplan and Yorke~\cite{Kaplan79a} as
part of a conjecture that related Lyapunov exponents to fractal dimension.
If $\lambda_1 \ge \lambda_2 \ge \cdots $ are
the ordered Lyapunov exponents of a dynamical
system, and $k$ is the largest integer such that
$\lambda_1 + \cdots + \lambda_{k} > 0$, then the Lyapunov dimension
is given by
\be
	D_\lambda = k + \frac{\lambda_1 + \cdots + \lambda_{k}}{|\lambda_{k+1}|}.
\ee
Note that the Lyapunov dimension depends only on the largest $k+1$
Lyapunov exponents.

	The effect of a general (causal~\cite{acausal-note})
IIR filter is to add new negative
Lyapunov exponents to the dynamics.  This is readily seen in the case
of the AR(1) filter.  As discussed by Badii\etal~\cite{Badii88},
the filter
\be
	e_t = b e_{t-1} + x_t
	\label{eq-ar-one}
\ee
adds a new variable ($e_t$) to the dynamical system, and a new Lyapunov
exponent $\lambda = \log |b|$.  A higher
order AR($q$) filter
\be
	e_t = \sum_{k=1}^q b_k e_{t-k} + x_t
	\label{eq-ar-q}
\ee
can be factored to give $q$ new variables, and $q$ new Lyapunov exponents.
If $z_1,\ldots,z_q$ denote the $q$ roots of the associated polynomial
$Q(z) = 1-\sum_{i=1}^q b_q z^q = (1-z/z_1)\cdots(1-z/z_q)$,
then we can write \eq{ar-q} as a system
of $q$ equations:
\bea
	e^{(1)}_t &=& (1/z_1) e^{(1)}_{t-1} + e^{(2)}_t \nonumber \\
	          &\vdots& \\
	e^{(q)}_t &=& (1/z_q) e^{(q)}_{t-1} + x_t \nonumber
\eea
and the new Lyapunov exponents are given by $\lambda_i = \log (|1/z_i|)
= -\log |z_i|$ for $i=1,\ldots,q$.  Note that the roots $z_i$ must all
lie outside the unit circle ($|z_i|>1$) for the filter in \eq{ar-q} to
be stable, and in this case all the new Lyapunov exponents are negative.

If we rewrite the AR($q$) filter above in terms of its equivalent
MA($\infty$) filter; that is,
\be
	e_t = \sum_{k=0}^\infty a_k x_{t-k},
	\label{eq-ma-infty}
\ee
then the polynomial
\be
	P(z) = \sum_{k=0}^\infty a_k z^k
	\label{eq-pz}
\ee
will satisfy
$P(z)=1/Q(z)$ and will have poles $z_1,\ldots,z_q$ where $Q(z)$ has
roots.

All of this is motivation for the following statements: The new
Lyapunov exponents generated by an IIR filter given in \eq{ma-infty}
are $\lambda_i = -\log|z_i|$ where $z_i$ are the poles of the
polynomial in \eq{pz}.  If the filter is invertible and has bounded
coefficients $a_k$, then there will be no poles or zeros inside the
unit circle.

	Now, we wish to consider the particular IIR filter that
corresponds to bleaching.  This is given by
\eq{bleach} with $q=\infty$, and has the property
\be
	\langle e_t e_{t-\tau} \rangle = \sigma^2\delta_{0,\tau}.
	\label{eq-etetau}
\ee
Let us write the causal inverse of the filter in
\eq{bleach} as
\be
	x_t = \sum_{k=0}^\infty b_k e_{t-k}
\ee
where the $b$'s are given as the coefficients of the polynomial
\be
	Q(z) = \sum_{k=0}^\infty b_k z^k = \frac{1}{P(z)}.
\ee
We can exploit the condition in \eq{etetau} by writing
\bea
	\langle x_t x_{t+k} \rangle &=& \sum_{i=0}^\infty
		\sum_{j=0}^\infty b_i b_j \langle e_{t-i} e_{t+k-j} \rangle
		\nonumber \\
	&=& \sigma^2 \sum_{i=0}^\infty b_i b_{i+k}.
\eea
If we introduce the autocorrelation
``generating function'' (Ref.~\cite{Anderson71}, Sec.~5.7.1)
which is the polynomial
\be
	{\cal A}(z) = \sum_{k=-\infty}^\infty A_k z^k
\ee
where $A_k$ is the autocorrelation function defined in \eq{autocorr},
it is not hard to show that
\be
	{\cal A}(z) = cQ(z)Q(z^{-1})
\ee
where $c$ is a constant multiplier.
Thus, if $z_o$ is a root of $Q(z)$, then both $z_o$ and $z_o^{-1}$ are
roots of ${\cal A}(z)$.
It follows that the largest new Lyapunov exponent introduced by the
bleaching is given by
\be
	\lambda_o = -\log |z_o|
	\label{eq-lambda-o}
\ee
where $z_o$ is the
smallest root of the autocorrelation generating function ${\cal A}(z)$ that is
{\em outside} the unit circle.

	Since the Lyapunov dimension $D_\lambda$ depends only on the
largest $D$ Lyapunov exponents, where $D=\lceil D_\lambda\rceil$, a
new Lyapunov exponent will change the Lyapunov dimension only if it is
larger than $\lambda_D$, the $D$th largest Lyapunov exponent of the original
dynamics.  These $D$ largest Lyapunov exponents are the only ones
accessible from a trajectory that is on the attractor;
\v{C}enys~\cite{Cenys93} calls these the ``internal'' Lyapunov
exponents.

	Therefore, a bleaching filter will change the Lyapunov dimension
whenever $\lambda_o$ of \eq{lambda-o} is greater than
the smallest accessible (or ``internal'') Lyapunov exponent $\lambda_D$.
One can think of this in terms of two time scales:
one is the ``linear'' timescale
assocated with the autocorrelation function, and the other is the
``nonlinear'' timescale associated with the smallest accessible Lyapunov
exponent.  When the linear timescale is longer than the nonlinear timescale,
then bleaching will, in the infinite data limit, actually change the
structure of the attractor.

We have already seen, however, that even finite-order bleaching can
have a dramatic effect on estimates of nonlinear invariants, and in the
following section we outline an approach for quantifying that
effect.

\subsection{FILTERING, EMBEDDING, AND PROJECTING}
\label{sec-embed}

As noted in Refs.~\cite{Sauer91,Broomhead92,Sauer93},
the issue of prefiltering can be recast as an embedding problem.
Given a time series $x_t$, one can ask which of the following
``embeddings'' best describes the actual state of the system at
time $t$:
\bea
	S^{(0)}_t &=& ( x_t, x_{t-1}, x_{t-2} )\mbox{; or}
	\label{eq-trunc-nofilter} \\
	S^{(a)}_t &=& ( e_t, e_{t-1}, e_{t-2} )\nonumber\\
	          &=& ( x_t - a x_{t-1},\,\, x_{t-1}-a x_{t-2},\,\, x_{t-2} - a
x_{t-3} ).
	\label{eq-trunc-wfilter}
\eea
And in fact, {\em both} of these are projections from the higher
dimensional space: $(x_t, x_{t-1}, x_{t-2}, x_{t-3})$.
Indeed, the two panels in \fig{henon-bleach} can be viewed
as two different projections from the eight-dimensional space
$(x_t, x_{t-1}, \ldots, x_{t-7})$.  Thus the twin issues of
optimal filtering and optimal embedding can both be rephrased
in terms of optimal projection.

There are a number of criteria for judging the quality of an
embedding.  Operational criteria would define the fitness of an
embedding in terms of how well it permits nonlinear forecasting or
dimension estimation.  More direct criteria have also been
proposed~\cite{Casdagli91c,Fraser86,Liebert89b,Aleksic91}.

In particular, the approach suggested by
Casdagli\etal~\cite{Casdagli91c} compares different embeddings
according to how measurement noise is amplified when the embedded
state is mapped back to the original state space.  The authors define
a ``distortion'' $\delta$ which is related to this amplification.  In
this section, we will measure $\delta$ for bleached and unbleached
data.  We will also introduce a new quantity, $\gamma$, which we will
call ``stretching;'' this measures how much a spherical
(infinitesimal) noise ball in the original state space will be
stretched in going to the embedded space.  This new quantity, though
also a local quantity (by which we mean it does not depend on global
information in the attractor, such as how the attractor is ``folded''
by the dynamics), provides complimentary information about the
embedding.

Following Casdagli\etal~\cite{Casdagli91c}, let $\Phi$ be the map
that takes the actual state into the time delay embedding:
$\Phi: {\bf R}^d \to {\bf R}^{m+q}$, where $d$ is the dimension of
the actual state space. Let $\Psi_{\rm T}$ and $\Psi_{\rm B}$ denote
two different projections of the time delay embedding from $m+q$
coordinates to $m$ coordinates.  $\Psi_{\rm T}$ is just the map that
projects out the first $m$ coordinates, and $\Psi_{\rm B}$ is
the projection that corresponds to bleaching the time series.
Let $D\Phi$ and $D\Psi$ be the Jacobians of these maps; in general
they will depend on the location in state space.

{}From $D\Phi$ and $D\Psi$, a ``distortion'' matrix can be
defined~\cite[Eq.~(75)]{Casdagli91c}
\def\tr{^{\mbox{\tiny T}}}
\be
	\Sigma = \left[ D\Phi\tr D\Psi\tr
	(D\Psi D\Psi\tr)^{-1}
	D\Psi D\Phi \right]^{-1}
\ee
and the distortion itself is given by $\delta = \sqrt{\mbox{Trace}(\Sigma)}$.
Casdagli\etal~\cite{Casdagli91c} have noted that if $\Psi$ is invertible,
there will be no effect at all on distortion.  However, even
if the {\em filter} is invertible, the matrix
$\Psi$ is still a projection, and it is {\em not} invertible.

We define the stretching matrix simply as the inverse of the distortion
matrix, and so the stretching itself is
$\gamma = \sqrt{\mbox{Trace}(\Sigma^{-1})}$.  Note that while the
distortion is sensitive to large eigenvalues of $\Sigma$, the
stretching is sensitive to small eigenvalues of $\Sigma$.  A more
comprehensive theory might consider the full eigenvalue spectrum.
Note also, in comparison with
Eq.~(83) of Ref.~\cite{Casdagli91c}, that this is a local quantity that
appears related to estimation error.

\ignore{Pecora has interpreted the theory of Casdagli\etal as
saying that the optimum filter is no filter at all --- his own
theory of filtering also has this unsatisfying feature.  But we should
emphasize that this is an incorrect interpretation.  The theory
says that the best projection of, say, $(x_t,\ldots,x_m)$ is
no projection at all.  But it does provide a comparison of
\eq{trunc-nofilter} and \eq{trunc-wfilter}, as two different projections
of ${\bf R}^4\to{\bf R}^3$, and does not necessarily choose the first
as optimal.}

\begin{figure}[htbp]
\centerline{\hbox{
\psfigure{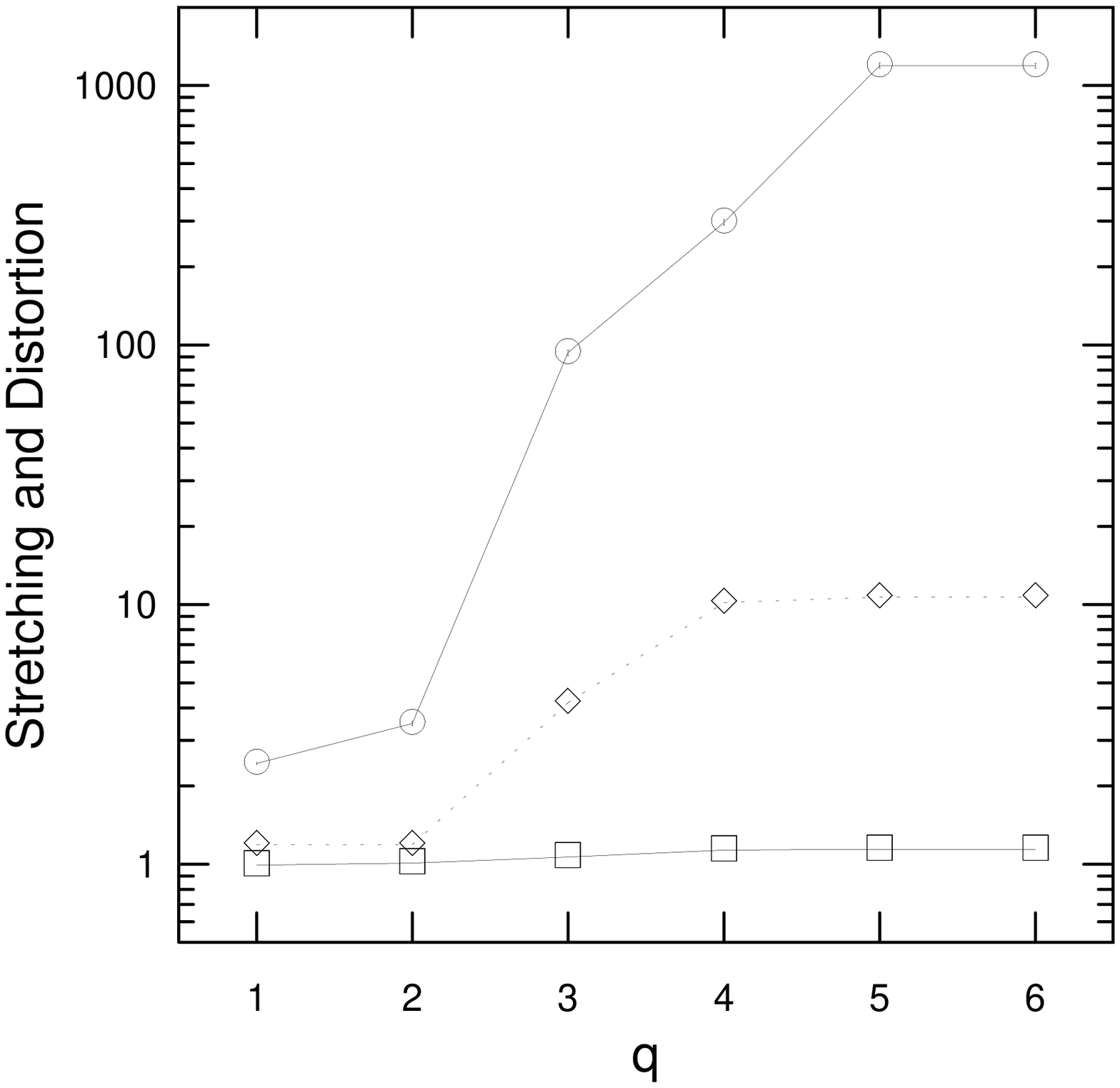}{\fullfigsize}
}}
\kaption{Mean ($\Box$) and maximum ($\Diamond$) distortion,
and mean stretching ($\circ$)
as a function of bleaching order $q$, for an
$m=3$ embedding of the H\'enon time series.
The effect of bleaching on distortion is quite small;
on average it is very near unity, and at the few points where
the effect is maximal, it is only of
order ten.  The average
stretching, by contrast (compare circles ($\circ$) with squares ($\Box$)),
increases dramatically with $q$.}
\label{fig-maxdistortion}
\end{figure}

In \ignore{\fig{distortion} and }\fig{maxdistortion}, we compare the
distortion for the embeddings of a H\'enon time series bleached at
increasing levels of $q$.  We again remark that an embedding dimension
of $m=3$ is sufficient for all finite values of $q$ because the
H\'enon attractor has a dimension $d\approx 1.3$, and
$m>2d$.  It appears from these figures that bleaching
does not induce considerable distortion, but that it does a phenomenal
amount of stretching.

\ignore{
\begin{figure}[htbp]
\centerline{\hbox{
\psfigure{del-aQ-m3.ps}{\fullfigsize}
}}
\kaption{Distortion of the $m=3$ residual embedding plotted against $x_t$
for various bleaching orders $q$, for a time series from a single
H\'enon map.\tcomment{Be sure to re-check those coefficients}
}
\label{fig-distortion}
\end{figure}
}

\begin{figure}[htbp]
\ \centerline{\hbox{
\psfigure{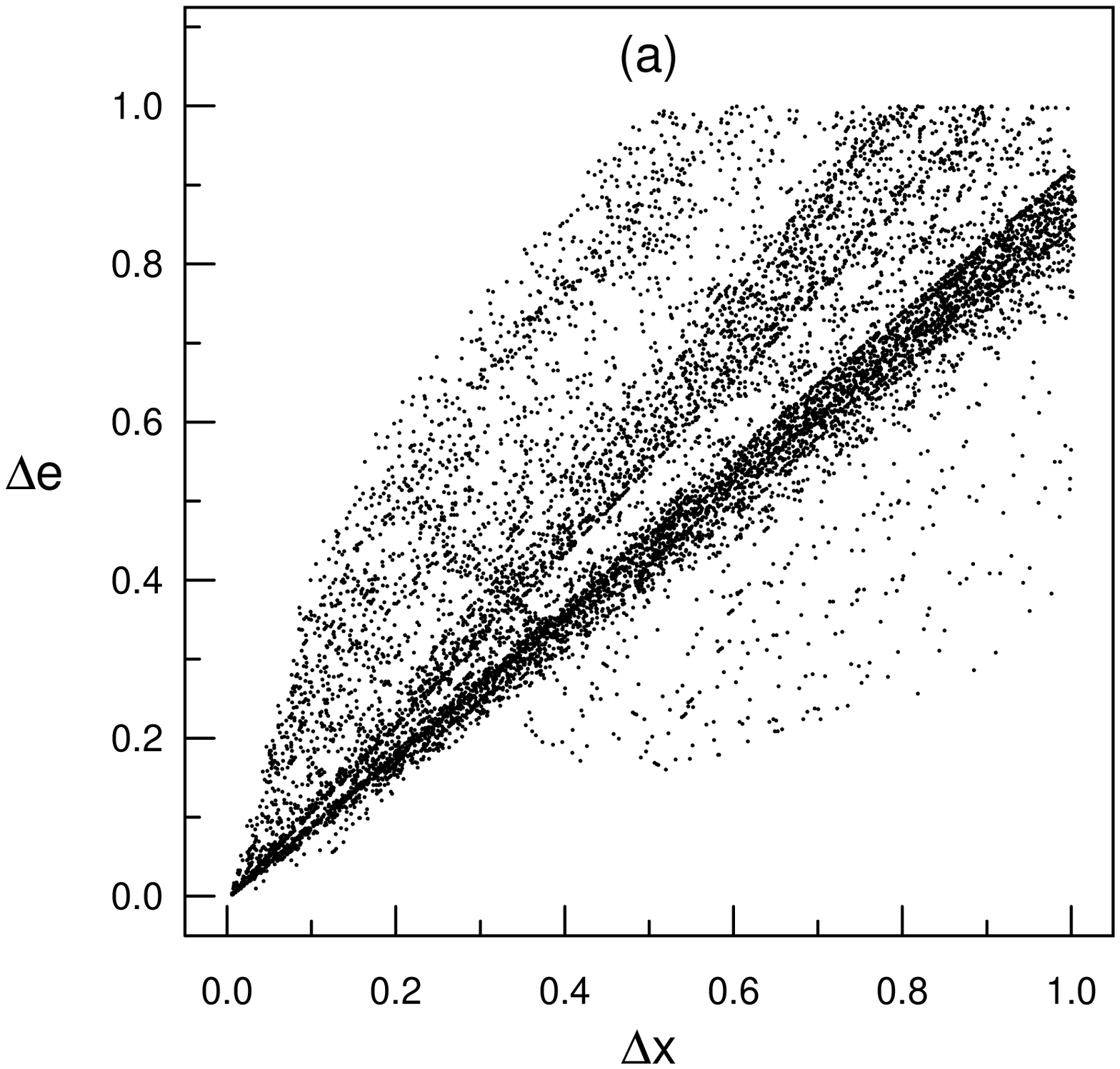}{\fullfigsize}
}}
\ifcompact\vspace{2ex}\fi
\ \centerline{\hbox{
\psfigure{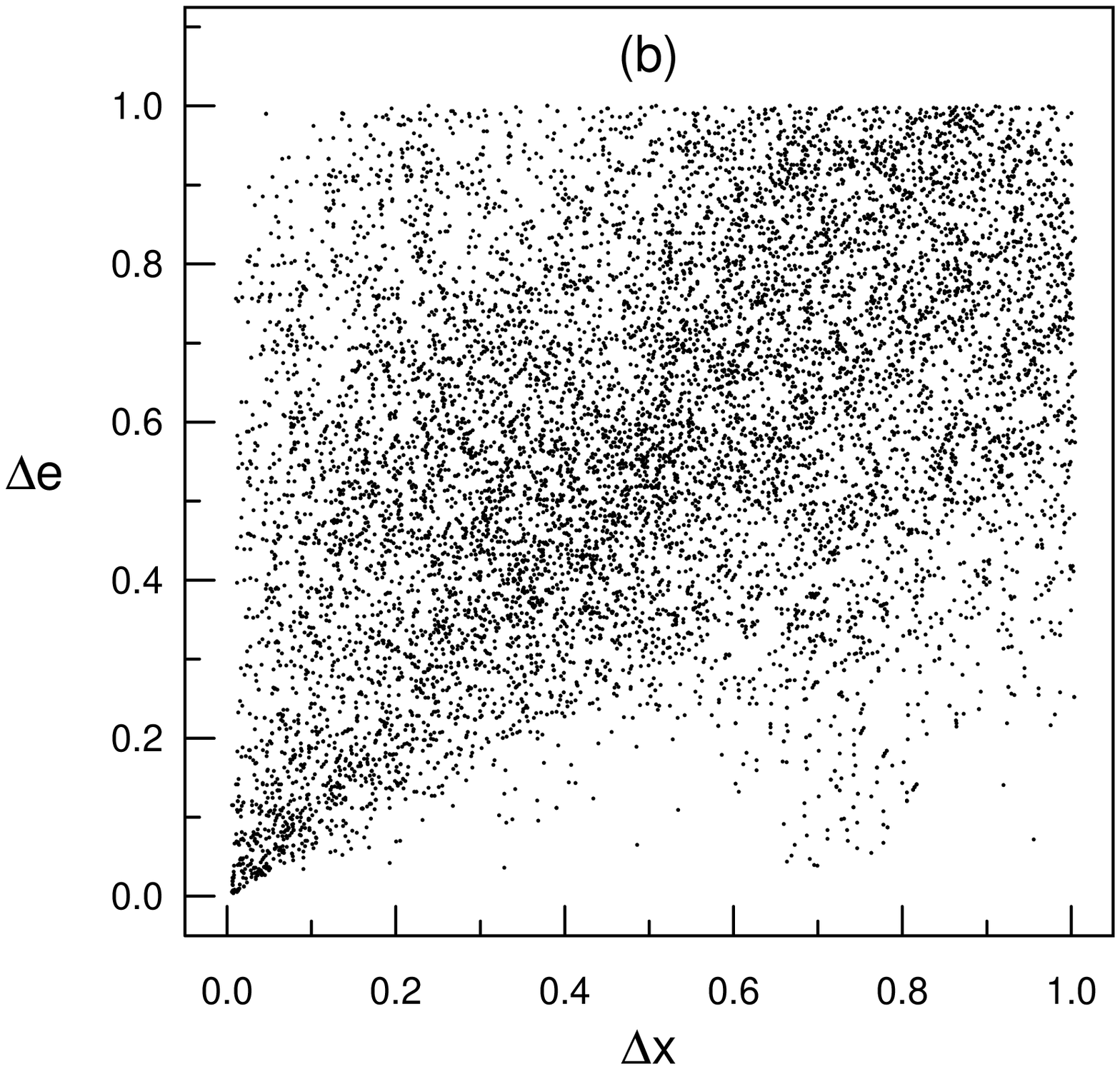}{\fullfigsize}
}}
\kaption{The distance between a pair of points in the residual embedding
($\Delta e$) is plotted against the distance between the same pair of points
in the original space
($\Delta x$).  Note that there are
many points for which the distance in the residual embedding space
is much larger than the corresponding distance in the original
embedding, {\em i.e.,} $\Delta e \gg \Delta x$.  There are relatively
fewer points for which $\Delta e \ll \Delta x$, which suggests that
stretching, and not distortion, is the dominant factor in this case.
The time series is
from the H\'enon map and the embedding dimension
is $m=3$.  {\bf (a)} Only a single bleaching term, $q=1$.
{\bf (b)} Full bleaching, $q=6$.}
\label{fig-de-vs-dx}
\end{figure}

Another way of looking at what is happening can be seen in
\fig{de-vs-dx}.  Here, distances between pairs of residuals ($\Delta e_{ij}
= | \vec{e}_i - \vec{e}_j |$) are plotted as a scatterplot against the
corresponding distances for the original time series ($\Delta x_{ij}
= | \vec{x}_i - \vec{x}_j |$).  Again, we use an embedding dimension
of m=3, so that the attractors do not overlap themselves.
  We see a large
population of points for which $\Delta e \gg \Delta x$; these are
pairs of points which are close in the original coordinates, but
have been stretched far apart in the residual coordinates.

\section{CONCLUDING REMARKS}

	In a variety of numerical experiments, we have described the ill
effects of bleaching on nonlinear models of chaotic time series data.
We have shown in particular that for detecting nonlinearity,
it is often better to compare the given time series with stochastic
data that mimics its autocorrelation than to try and subtract out the
autocorrelation altogether.  This led us to suggest modifications to some
standard residual-based statistics, among them the BDS and
the McLeod-Li statistics.  From the point of view of model building, we
have seen that fitting of residuals can cost several orders of
magnitude in accuracy of fit, compared to fitting the original data.
On the other hand, having demonstrated cases where linear
prefiltering is disadvantageous, we have also seen cases where {\em
some} linear filtering helps.

We have also done experiments with
the correlation dimension, and while these results are not shown
(but see Sauer and Yorke~\cite{Sauer93} for a demonstration of how
linear filtering can affect estimates of correlation dimension), these
estimates are also seriously degraded by the effects of bleaching.
Although we have not done the relevant numerical experiments, we suspect
that indiscriminate bleaching will have a similarly deleterious effect
on estimates of Lyapunov exponent, or upon the tests for determinism
advocated by Casdagli~\cite{Casdagli91e} and Kaplan~\cite{Kaplan92}.

Brock~\cite{Brock86c} has noted that residual-based statistics ``may
mis-identify deterministic chaos as random noise in a short data
set,'' but chose to use a residual-based statistic in his study for
reasons that were to some extent motivated by the considerable
interest at the time in AR(2) models with roots near the unit
circle~\cite{Brock-historical}; for such systems, as we noted in
\sect{unfilter}, pre-whitening can help.
More recent modeling by Brock~{\it et al.}~\cite{Brock91c} used a
direct resampling (surrogate data) method for rejecting a variety of
null hypotheses.  While these conclusions suggest that earlier
failures to detect nonlinearity in various economic time series may be
vulnerable to more powerful tests that are not based on residuals, we
consider this unlikely, because our tests are more powerful when the
alternative hypothesis is chaos, and we have seen no convincing
evidence of chaos in financial time series.  On the other hand, we
{\em are} saying that if chaos is the alternative hypothesis, then
residual based statistics are probably not as powerful as direct comparisons
with similarly autocorrelated (surrogate) data.

\ignore{There is also the notion of `optimal' tests which
test a given null hypothesis against a specific alternative.
While choosing a specific alternative may not be useful in
itself, because one would like to have power over a wide range
of tests, one does have chaos as an alternative inthe back of
ones mind, that is a certain class of alternatives.  So,
ideally, we'd like to have not just linear vs. nonlinear but,
somehow linear vs. chaos.  Basically you interpret the results
the same, but the statistic particularly adept at picking out
chaos.  Like the asymmetry statistic, it does distinguish linear
from nonlinear, but in particular, it distinguishes reversible
from non-reversible.}

Scargle~\cite{Scargle89} has suggested that a kind of nonlinear Wold
decomposition theorem can be derived in which the chaotic process is
rewritten as a linear filter of ``white chaos.''  This uncorrelated
process is just the residual time series $e_t$ of \eq{bleach}, and our
main point in this \whatthisis\ is that the residual time series can
be much more complicated and difficult to work with than the raw time
series.  White chaos pays a price for its whiteness.  Actually, the
algorithm Scargle used for determining the chaotic innovation was more
complicated than that of \eq{bleach}, and in a later
paper~\cite{Scargle90}, he recognizes that this algorithm does not in
general produce a time series that is in fact uncorrelated.  We do not
know if the effective prefilter that Scargle ultimately proposes is
in general beneficial or detrimental to nonlinear modeling of the time
series.

Sugihara and May~\cite{Sugihara90} have noted that their test for
chaos based on prediction error can be fooled by autocorrelated noise,
and they suggest first-differencing as a method of removing
autocorrelation.  Although this may be useful in some cases, we argue
that this general approach is likely to be problematic on several
counts.  One, first differencing does not necessarily remove
autocorrelation, and in some cases can enhance it; two, in cases where
autocorrelation is not removed, the test is still vulnerable to linear
artifacts; and three, even if the autocorrelation is significantly
removed, the state space structure can become significantly distorted,
and the power of the test for detecting nonlinearity (let alone chaos)
will have been compromised.

\appendix
%
\ignore{

PROOF that the best linear fit leads to uncorrelated residuals

}

\section*{APPENDIX: DEMONSTRATION THAT BEST FIT RESIDUALS ARE WHITE}

In this appendix, we show that in the limit $q\to\infty$,
the best linear fit leads to
uncorrelated residuals.

Let $\hat x_t$ be the ``best'' linear estimator for $x_t$,
in the sense of minimizing the variance $\langle e_t^2\rangle$
of the residuals, where $e_t = x_t - \hat x_t$.
Consider an arbitrary time delay $\tau>0$, and let
\be
	\lambda = \frac{\langle e_t e_{t-\tau}\rangle}{\langle e_t^2\rangle}.
\ee
Since we want to show that the residuals
are uncorrelated, what we want to show is that $\lambda=0$.
Our approach will be to show is that if $\lambda\ne0$, then a better
linear estimator than $\hat x_t$ can be constructed, contradicting
the hypothesis that $\hat x_t$ was optimal.

Begin by noting that a good estimator for $e_t$ is given by
\be
	\hat e_t = \lambda e_{t-\tau}
\ee
so that a new linear estimator for $x_t$ can be defined by
\bea
	\hat{\hat x}_t &=& \hat x_t + \hat e_t \nonumber \\ &=&
	\hat x_t + \lambda(x_{t-\tau} - \hat x_{t-\tau}).
\eea
Note that this too is an ordinary linear estimator for $x_t$
in terms of past values $(x_{t-1},\ldots)$.  Note also, that
if $\hat x$ were restricted to finite order $q$, then $\hat{\hat x}$
would be of order $\tau+q$, so this argument does not apply
to finite estimators, except through a separate result which we will
not show here (see, for
instance, Theorem 7.6.6 in Anderson~\cite{Anderson71})
that finite-order estimators approximate
infinite-order estimators as $q\to\infty$.

Let $\epsilon_t$ be the residuals from this new estimator:
$\epsilon_t = x_t - \hat{\hat x}_t$.  Then
\bea
	\langle \epsilon_t^2 \rangle &=&
	\langle (x_t - \hat{\hat x}_t)^2 \rangle \nonumber \\ &=&
	\langle (x_t - [\hat x_t + \hat e_t])^2 \rangle \nonumber \\ &=&
	\langle ([x_t - \hat x_t] - \hat e_t )^2 \rangle \nonumber \\ &=&
	\langle (e_t - \lambda e_{t-\tau})^2 \rangle \nonumber \\ &=&
	\langle e_t^2 \rangle - 2\lambda\langle e_t e_{t-\tau}\rangle +
	\lambda^2\langle e_{t-\tau}^2 \rangle \nonumber \\ &=&
	(1-\lambda^2)\langle e_t^2 \rangle.
\eea
{}From our original hypothesis that $\hat x$ was the optimum linear
predictor, we have $\langle e_t^2\rangle \le \langle \epsilon_t^2\rangle$,
which requires $\lambda=0$, and implies that $\langle e_t e_{t-\tau} \rangle$
is zero for all $\tau > 0$.  That is, the residuals have no autocorrelation;
they are white.

\section*{ACKNOWLEDGEMENTS}

We are pleased to acknowledge Bryan Galdrikian, Andr\'e Longtin, and
Doyne Farmer, who collaborated with us in developing the method of
surrogate data.  We are also grateful to Blake LeBaron, William Brock,
Tim Sauer, and Lou Pecora for many useful discussions.  This work was
partially supported by the National Institute for Mental Health under
Grant No. 1-R01-MH47184, and performed under the auspices of the
U.S. Department of Energy.


{
\ifcompact\footnotesize\fi

\begin{thebibliography}{10}
\bibitem{choice-of-mcleod-li}
We chose this version of McLeod-Li as a representative of the ``conventional''
  approach to detecting nonlinearity because it is simple to understand and
  implement; there are better approaches (see Tong~\cite{Tong90}) but these
  tend to involve massive regressions. Our purpose here is meant to be more
  illustrative than comparative.
\bibitem{acausal-note}
The effects of acausal filters are discussed by Mitschke~\cite{Mitschke90} and
  by Pecora and Carroll~\cite{Pecora93}.
\end{thebibliography}

}

\ifnotcompact
	\newpage
	\renewcommand{\contentsline}[3]{#2}
	\renewcommand{\numberline}[2]{{\bf Fig.~#1.} {#2}\\[2ex]}
	\listoffigures
\fi


\end{document}